\documentclass[12pt]{article}
\usepackage{amsmath}
\usepackage{amsfonts}
\usepackage{epsfig}
\usepackage{amssymb}
\usepackage{comment}
\usepackage{a4wide}
\usepackage{color}
\usepackage{float}
\usepackage{times}[20]
\usepackage{mathptmx}
\newtheorem{thm}{Theorem}[section]

\newtheorem{lemma}{Lemma}[section]
\newtheorem{theorem}{Theorem}[section]
\newcommand{\qed}{\hfill \rule {1ex}{1ex}}

\begin{document}

\title{The Hubbard model at half-filling, part III~:  the  lower bound on the
  self-energy}

\author{St\'ephane Afchain, Jacques Magnen, \\  Centre de Physique Th\'eorique, CNRS, UMR 7644, \'Ecole Polytechnique, \\ 91128 Palaiseau - France \\ Vincent Rivasseau \\ Laboratoire de Physique Th\'eorique, CNRS, UMR 8627\\
Universit\'e de Paris-Sud, 91405 Orsay}

\maketitle

\begin{abstract} We complete the proof  that the two-dimensional Hubbard model  at half-filling is {\it not} a 
Fermi liquid in the mathematically precise sense of Salmhofer, by establishing a lower 
 bound on a second derivative in momentum of the first non-trivial self-energy graph.
\end{abstract}

\section{Introduction}

This paper is the third of a series (\cite{R1}-\cite{AMR}) devoted to the rigorous mathematical study of the two-dimensional Hubbard model at half-filling above the transition temperature to the expected low temperature region, which becomes N\'eel-ordered at zero temperature. 
The goal of this series was to prove that this model does not obey Salmhofer's criterion for Fermi liquid behavior of interacting Fermion systems at equilibrium (\cite{S1}-\cite{Salm}). In this way, this model differs sharply from those with a Fermi surface close to the circle, which obey Salmhofer's criterion (\cite{DR1}-\cite{DR2}-\cite{BGM}). 

In the first paper \cite{R1} the convergent contributions of the model were
bounded in the domain $| \lambda | \log^2 T \leq K$. In the second one
\cite{AMR}, renormalization of the self-energy was performed to complete the
proof of analyticity in the coupling constant of all the correlation functions
in that domain. Salmhofer's criterion requires beyond this analyticity that
the self-energy (in momentum space) is uniformly bounded together with its
first and second derivatives  in a  domain $| \lambda |  |\log T| \leq K$.  In
this paper we prove that a certain second derivative of the self-energy at a
particular value of the external momentum is \textit{not} uniformly bounded in
the domain $| \lambda | \log^2 T \leq K$ where we have established analyticity. This domain being smaller than the Salmhofer's one, it completes the proof that the two-dimensional half-filled Hubbard model is not a Fermi liquid. In conclusion, when we move from low filling to half-filling, the Hubbard model must undergo a cross-over from Fermi to non-Fermi (in fact Luttinger) liquid behavior. This solves the controversy on the nature of two-dimensional Fermionic systems in their ordinary phase \cite{And}. We refer to \cite{R1}-\cite{AMR}-\cite{Salm} for a more complete review and further references on mathematical study of interacting Fermions.

\section{Recall of notations}

The two-dimensional Hubbard model is defined on the lattice
$\mathbb{Z}^2$. Fixing a temperature $T > 0$, the "imaginary time", denoted
$x_0$, belongs to the real interval $\left[ -\frac{1}{T}, \ \frac{1}{T}
\right[$. In the following, we shall denote $\beta = \frac{1}{T}$. Indeed this
interval should be thought of as a circle of length $2 \beta$, that is
$\mathbb{R} / 2\beta \mathbb{Z}$. Consequently, the momentum space, which is the dual of $\mathbb{R} / 2 \beta \mathbb{Z} \times \mathbb{Z}^2$ in the sense of the Fourier transform, is $\pi T \mathbb{Z} \times [\mathbb{R} / 2\pi \mathbb{Z}]^2$. The torus $[\mathbb{R} / 2\pi \mathbb{Z}]^2$ will be represented by the square $[-\pi, \ \pi[^2$, with periodic boundary conditions. 

In Fourier variables, the expression of the propagator at half-filling reads~:
\begin{equation}
C(k_0, k_1, k_2) = \frac{1}{i k_0 - \cos k_1 - \cos k_2}
\end{equation}
if $k_0 = (2n+1) \pi T$ for some $n \in \mathbb{Z}$. If $k_0 = 2n \pi T$, $C(k_0, k_1, k_2) = 0$ because in the formalism of Fermionic theories at finite temperature, the propagator has an antiperiod $\beta$ with respect to the $x_0$ variable and therefore each Fourier coefficient of even order vanishes. With a slight abuse of language, we can say that $C(k_0,k_1,k_2)$ is only defined for $k_0 = (2n + 1) \pi T$. This set of values is called the Matsubara frequencies. 

The expression of the propagator in real space is deduced by Fourier transform~:
\begin{equation}
C(x_0, x_1, x_2) = \frac{1}{(2\pi)^3}\int dk_0 \int dk_1 \int dk_2 \ \frac{e^{i k.x} }{i k_0 - \cos k_1 - \cos k_2}
\end{equation}
where we adopt the notations of \cite{R1}, namely the integral $\int dk_0$ really means the discrete sum over the Matsubara frequencies $2\pi T \sum_{n \in \mathbb{Z}} \eta ((2n + 1) \pi T)$ (with $k_0 = (2n + 1) \pi T$), whereas the integrals over $k_1$ and $k_2$ are "true" integrals, for $(k_1, k_2) \in [-\pi, \ \pi[^2$. We have added an ultraviolet cutoff $\eta (k_0)$, which is a fixed $C_0^{\infty}$ (which e.g. is 1
for $0 \le k_0 \le 1$ and 0 for $0 \le k_0 \ge 2$) in order to avoid some
technicalities irrelevant for our main result, namely the fact that the integrand without this cutoff is not absolutely summable with
respect to $k_0$ or $n$.

\begin{figure}[H]
\centerline{\resizebox{6cm}{!}{\input{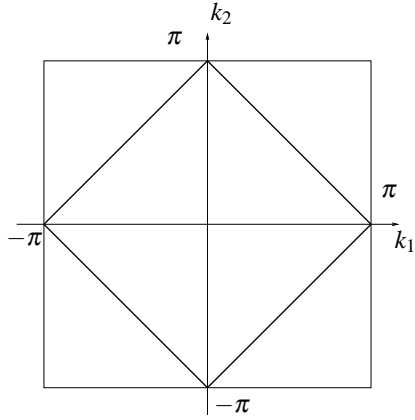}}}
\label{surface_de_Fermi}
\caption{The square $[-\pi , \pi[^2$ and the Fermi surface}
\end{figure}

For our analysis, it will be convenient to introduce another parametrization
of the spaces $[-\pi, \pi[^2$ and $\mathbb{Z}^2$. The idea is to "rotate" the
Fermi surface of Figure \ref{surface_de_Fermi} by an angle of $\frac{\pi}{4}$. In the $k_0 = 0$ plane, it is defined by $\cos k_1 + \cos k_2 = 0$, which is equivalent to $k_2 = \pi \pm k_1$ or $k_2 = - \pi \pm k_1$.

\begin{figure}[H]
\centerline{\resizebox{8cm}{!}{\input{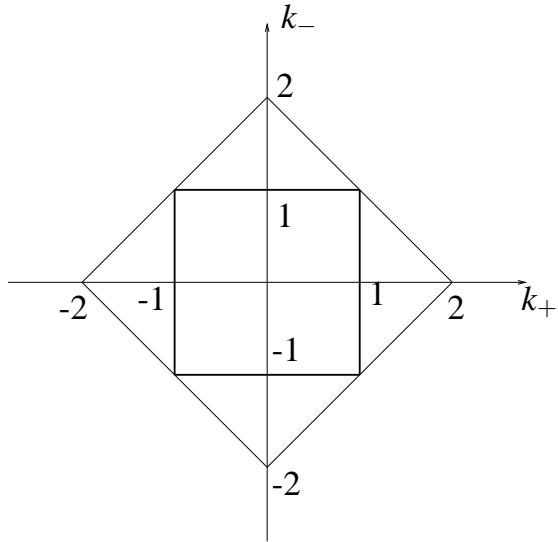}}}
\caption{The domain of integration in $(k_+ , k_-)$ and the Fermi surface}
\label{surface_de_Fermi_tournée}
\end{figure}

Introducing the variables $k_{\pm} = \frac{k_1 \pm k_2}{\pi}
\Longleftrightarrow \begin{cases} k_1= \frac{\pi}{2} (k_+ + k_-) \\ k_2=
 \frac{\pi}{2} (k_+ - k_-) \end{cases} \hspace{-.3cm} $, the domain of 
  integration $(k_1, k_2) \in [-\pi, \pi[^2$ becomes the set~:
\begin{equation}
\mathcal{D} = \left\{  (k_+, k_-) \in [-2, 2]^2 \ \text{with} \ \begin{cases}-2 \leq k_+ \leq 0 \\ -2 - k_+ \leq k_- \leq 2+ k_+ \end{cases} \hspace{-.5cm}\text{or} \ \begin{cases} 0 \leq k_+ \leq 2 \\ -2 + k_+ \leq k_- \leq 2 - k_+ \end{cases}\right\} \; .
\end{equation}
As $\cos k_1 + \cos k_2 = 2 \cos  \frac{\pi}{2} k_+   \cos  \frac{\pi}{2} k_- $, the Fermi surface in the variables $k_\pm$ is simply defined by $k_+  = \pm 1 \ , k_-  = \pm 1 $. The new domain of integration, with the Fermi surface is represented on Figure \ref{surface_de_Fermi_tournée}.

In a dual way, we introduce new variables in real space, $x_+$ and $x_-$ in such a way that $k_1 x_1 + k_2 x_2 = k_+ x_+ + k_- x_-$. We have~:
\begin{equation}
\begin{cases}
x_+ =  \frac{\pi}{2} (x_1 + x_2) \\
x_- =  \frac{\pi}{2}  (x_1 - x_2) \ .
\end{cases}
\end{equation}
Observe that the image of the lattice $\mathbb{Z}^2$ by this change of
variable is not $\frac{\pi}{2} \mathbb{Z}^2$ but the subset 
\begin{equation}
\label{parity_condition}
S = \left\{  \left( \frac{\pi}{2} m, \frac{\pi}{2} n  \right), \ (m,n) \in
  \mathbb{Z}^2, \ m \equiv n [2] \right\} \ .
\end{equation}
In other words, the integers $m$ and $n$ must have same parity.

As the Jacobian of the transformation $\begin{pmatrix} k_1 \\ k_2 \end{pmatrix} = \begin{pmatrix} \frac{\pi}{2} & \frac{\pi}{2} \\ \frac{\pi}{2} & - \frac{\pi}{2} \end{pmatrix} \begin{pmatrix} k_+ \\ k_-  \end{pmatrix}$ is $J = - \frac{\pi^2}{2}$, we have~:
\begin{equation}
\int_{[-\pi, \pi]^2} dk_1 dk_2 \ \frac{e^{i(k_1 x_1 + k_2 x_2)}}{i k_0 - \cos
  k_1 - \cos k_2} = \frac{\pi^2}{2} \int_{\mathcal{D}} dk_+ dk_- \ \frac{e^{i
    (k_+ x_+ + k_- x_-)}}{i k_0 - 2 \cos  \frac{\pi}{2} k_+  \cos
  \frac{\pi}{2} k_-  } \ .
\end{equation}

But the domain $\mathcal{D}$ is not very convenient for practical computations, and therefore we would like the $k_+ k_-$ integration domain to factorize. Since the complement set $[-2, 2[^2 \backslash \mathcal{D}$ is another fundamental domain for the torus $\mathbb{R}^2 / 2\pi \mathbb{Z}^2$, we have~:
\begin{equation}
\int_{\mathcal{D}} dk_+ dk_- \ \frac{e^{i (k_+ x_+ + k_- x_-)}}{i k_0 - 2 \cos
  \frac{\pi}{2} k_+  \cos  \frac{\pi}{2} k_- } = \int_{[-2, 2[^2 \backslash
  \mathcal{D}} dk_+ dk_- \ \frac{e^{i (k_+ x_+ + k_- x_-)}}{i k_0 - 2 \cos
  \frac{\pi}{2} k_+  \cos  \frac{\pi}{2} k_- } \ .
\end{equation}
Hence~:
\begin{equation}
\int_{\mathcal{D}} dk_+ dk_- \ 
\frac{e^{i (k_+ x_+ + k_- x_-)}}{i k_0 - 2 \cos
  \frac{\pi}{2} k_+  \cos \frac{\pi}{2} k_- } = \frac{1}{2} 
\int_{[-2,2]^2} dk_+ dk_- \ \frac{e^{i (k_+ x_+ + k_- x_-)}}
{i k_0 - 2 \cos \frac{\pi}{2} k_+  \cos  \frac{\pi}{2} k_- } \ .
\end{equation}

Recapitulating, the expression of the propagator that we we take as our starting point is:
\begin{equation}
C(x_0, x_+, x_-) =   \int d^3 k  
\frac{e^{i (k_0 x_0 + k_+ x_+ + k_- x_-)}}{i k_0 
- 2 \cos \frac{\pi}{2} k_+  \cos  \frac{\pi}{2} k_- }
\label{propafina}
\end{equation}
for $x_\pm$ satisfying the parity condition (\ref{parity_condition}).
In \ref{propafina} the notation $\int d^3 k$ 
means 
\begin{equation}
\frac{1}{32 \pi} \int dk_0 \int_{[-2, 2]^2} dk_+dk_- \ ,
\end{equation}
where we recall that $\int dk_0$ means 
$2\pi T \sum_{n \in \mathbb{Z}} \eta((2n+1)\pi T)  $), since $k_0=(2n+1)\pi T$.

\begin{figure}[H]
\centerline{\resizebox{6cm}{!}{\input{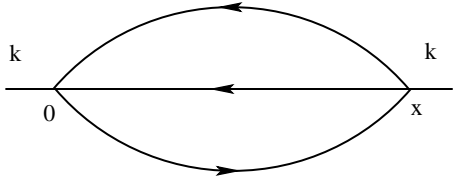}}}
\caption{The first non-trivial graph contributing to the self-energy}\label{graphe}
\end{figure}

Now, let us consider, in Fourier space, the amplitude of the graph $G$
represented on Figure \ref{graphe}, with an incoming momentum $k = (k_0, k_+, k_-)$. This amplitude is denoted $ A_G (k)$ and written
explicitly as $A_G (k_0, k_+, k_-) = \int d^3x \ C(x) 
\bar C(x)^2 e^{-i k.x}$ (where arrows join antifields to fields). 

More precisely, we shall consider the second momentum derivative in the $+$ direction of this quantity~:
\begin{equation}
\partial_+^2 A_G (k) = \int d^3x \ x_+^2 C(x) \bar{C(x)}^2 e^{-i k.x} \ .
\end{equation}

The quantity we are going to study is explicitly written~:
\begin{multline}
\partial_{+}^{2} A_G (\pi
T,1,0) = \int d^3 x \  x_+^2 \int d^3k_1
\frac{e^{ik_1.x}}
     {ik_{1,0}-2\cos \frac{\pi}{2}k_{1,+}\cos
       \frac{\pi}{2}k_{1,-}}  \\
\int d^3k_2
\frac{e^{ik_2.x}}
      {-ik_{2,0}-2\cos \frac{\pi}{2}k_{2,+}\cos
        \frac{\pi}{2}k_{2,-}} 
\int d^3k_3
\frac{e^{ik_3.x}}
{-ik_{3,0}-2\cos \frac{\pi}{2}k_{3,+}\cos
\frac{\pi}{2}k_{3,-}} \ 
e^{i(\pi T x_0+x_+)} \ ,
\end{multline}
where again $\int d^3 x$ includes the parity condition (\ref{parity_condition}).
We state now the main result of this paper~:

\begin{thm}
There exists some strictly positive constant $K$ such that, for $T$ small enough~:
\begin{equation}
\left| \partial_+^2 A_G (\pi T, 1, 0) \right| \geq \frac{K}{T} \ .
\end{equation}
\label{main_theorem}
\end{thm}
We recall that this result, joined to the analysis of \cite{AMR}, leads to the result that the self-energy of the model is not uniformly $\mathcal{C}^2$ in the domain $| \lambda | \log^2 T < K$ and therefore that the two-dimensional Hubbard model at half-filling is not a Fermi liquid.

\section{Plan of the proof}

Theorem (\ref{main_theorem}) will be proven thanks to a sequence of
lemmas. But before presenting these lemmas, let us give an overview
of our strategy. We use the sector decomposition introduced in \cite{R1} to
write~:
\begin{equation}
\label{sector_decomposition}
\partial_+^2 A_G (\pi T, 1, 0) = \sum_{\sigma_1, \sigma_2, \sigma_3} \int d^3x
\ x_+^2 C_{\sigma_1} (x) \bar{C}_{\sigma_2} (x) \bar{C}_{\sigma_3} (x) e^{- i(\pi T x_0 + x_+)} \ ,
\end{equation}
where a sector $\sigma$ is a triplet $(i , s_+ , s_-)$ with $0 \leq s_\pm \leq
i$ and $s_+ +s_- \leq i$.

The main idea is that in the sum over sectors of equation (\ref{sector_decomposition}), the
leading contribution is given by a restricted sum corresponding to sectors
close to the "vertical part" of the Fermi surface, defined by $k_+ = \pm
1$. To express this more precisely, let $\Lambda$ be an integer (whose value will be chosen later), which will play the role of a cut-off for the
sectors. We want to prove that as soon as one sector is not close
to $k_+ = \pm 1$, then we have a small contribution. Let us denote
$\sum_{\{i_j \}, \{s_j^+  \}, \{ s_j^- \}}^\Lambda$ the sum in which at least one sector is "far" from the vertical sides of the Fermi surface. Precisely, this means that at least one index $s_j^+$ is smaller than $i_{max} (T) - \Lambda$, where, as in \cite{R1}, $M^{- i_{max} (T)} \approx T$. This constrained sum can be written explicitly~:
\begin{multline}
\sum_{\{i_j  \}, \{ s_j^+ \}, \{s_j^-  \}}^\Lambda =  \sum_{i_1  , 
 s_1^- , \sigma_2, \sigma_3} \sum_{s_1^+ = 0}^{\inf (i_1, i_{max} (T) 
 -\Lambda} + \sum_{i_1  , 
  s_1^- , i_2, s_2^- , \sigma_3} \ 
\sum_{s_1^+ = i_{max} (T) -\Lambda }^{i_1}   \sum_{s_2^+ = 0}^{\inf (i_2, 
i_{max} (T) - \Lambda} \\ \hskip-1cm
+ \sum_{i_1  , s_1^- , i_2, s_2^- , i_3 , s_3^-} \ \sum_{s_1^+ = i_{max} (T) -\Lambda}^{i_1}  
\ \sum_{s_2^+ = i_{max} (T) -\Lambda }^{i_2}   
\sum_{s_3^+ =0}^{\inf (i_3, i_{max} (T) -
  \Lambda} \ .
\end{multline}
Defining~:
\begin{equation}
A_G^{\Lambda} (\pi T, 1, 0) = \sum_{\{ i_j \}, \{ s_j^+ \}, \{ s_j^-
\}}^{\Lambda} \int d^3 x \ C_{\sigma_1} (x) \bar{C}_{\sigma_2} (x)
\bar{C}_{\sigma_3} (x) e^{- i (\pi T x_0 + x_+)} \ , 
\end{equation}
we write~:
\begin{equation}
\partial_+^2 A_G (\pi T, 1, 0) = \partial_+^2 A_{G,\Lambda} (\pi T , 1 , 0) + \partial_+^2 A_G^{\Lambda} (\pi T, 1, 0)
\end{equation}
where $\partial_+^2 A_{G,\Lambda} (\pi T , 1 , 0) = \partial_+^2 A_G (\pi T, 1, 0) - \partial_+^2 A_G^{\Lambda} (\pi T, 1, 0)$ is expressed as a sum over sectors that are all close to $k_+ =
\pm 1$, i.e. such that each $s_j^+$ index is greater than $i_{max} (T) - \Lambda$.

Each sector appearing in the sum expressing $\partial_+^2 A_{G,\Lambda} (\pi T, 1, 0)$ will be divided into two disjoint subsectors, according to the sign of $\cos \frac{\pi}{2} k_+$. We recall that in \cite{R1}, the sectors
were defined as~:
\begin{equation}
\left| i k_0 - 2 \cos \frac{\pi}{2} k_{+} \cos \frac{\pi}{2} k_- \right|
\approx M^{- i} \ ,\quad \left| \cos \frac{\pi}{2} k_+ \right| \approx M^{-
s^+} \ , \quad \left| \cos \frac{\pi}{2} k_- \right| \approx M^{- s^-} \  .
\end{equation}

We shall call $\sigma^r$ and $\sigma^l$ ("right"
and "left") the subdomains of $\sigma$ corresponding to $\cos \frac{\pi}{2}
k_+ < 0$ and $\cos \frac{\pi}{2} k_+ > 0$ respectively. The underlying
motivation is that, if a momentum, say $k_1$, is close to the side $k_+ = 1$, by momentum conservation at each vertex, the other ones are 
necessarily close to the other side $k_+ = -1$. Let us state precisely this point~:

\begin{lemma}
In the sum expressing $\partial_+^2 A_{G,\Lambda} (\pi T, 1, 0)$, if one
sector is of the right type, then the other ones are of the left type.
\end{lemma} 

The proof is obvious by momentum conservation in the $+$ direction. We conclude that~:
\begin{multline}
\partial_+^2 A_{G,\Lambda}(\pi T, 1, 0) 
= \sum_{{\{ \sigma_j \}, i_j, s_j^+ >
i_{max} (T) - \Lambda } \atop{\sigma_1 \ \text{right}}} \int d^3x \ x_+^2
C_{\sigma_1} (x) \bar{C}_{\sigma_2} (x) \bar{C}_{\sigma_3} (x) e^{- i 
(\pi T x_0 + x_+)} \\
+ \sum_{\{ \sigma_j \}, i_j, s_j^+ > i_{max} (T) - \Lambda \atop \sigma_2 \
\text{right}} \int d^3x \ x_+^2
C_{\sigma_1} (x) \bar{C}_{\sigma_2} (x) \bar{C}_{\sigma_3} (x) e^{- i 
(\pi T  x_0 + x_+)} \\
+ \sum_{\{
\sigma_j \}, i_j, s_j^+ > i_{max} (T) - \Lambda \atop \sigma_3 \
\text{right}} \int d^3x \ x_+^2
C_{\sigma_1} (x) \bar{C}_{\sigma_2} (x) \bar{C}_{\sigma_3} (x) 
e^{-i(\pi T x_0 + x_+)} \ .
\end{multline}
Among these three contributions, the last two ones are indeed equal, and we
have~:
\begin{multline} \label{two_contributions}
\partial_+^2 A_{G,\Lambda} 
(\pi T, 1, 0) = \sum_{\{ \sigma_j \}, i_j, s_j^+ >
 i_{max} (T) - \Lambda \atop \sigma_1 \ \text{right}} \int d^3x \ x_+^2
C_{\sigma_1} (x) \bar{C}_{\sigma_2} (x) \bar{C}_{\sigma_3} (x) 
e^{-i(\pi T x_0 + x_+)} \\
+ 2 \sum_{\{ \sigma_j\}, i_j, s_j^+ > i_{max} (T) - \Lambda \atop \sigma_2 \ \text{right}}
\int d^3x \ x_+^2 C_{\sigma_1} (x) \bar{C}_{\sigma_2} (x) \bar{C}_{\sigma_3} (x)  e^{-i(\pi T x_0 + x_+)} \ .
\end{multline}

In each sum, we replace the $\cos \frac{\pi}{2} k_+$ appearing in the
propagators by their Taylor expansions in the neighborhood of $+1$ in a right
sector, and in a neighborhood of $-1$ in a left sector. We have $\cos
\frac{\pi}{2} k_+ \approx - \frac{\pi}{2} (k_+ - 1)$ for $k_+$ in the
neighborhood of 1, in which case we put 
$q_+ = (k_+ - 1)$ and $\cos \frac{\pi}{2} k_+ \approx \frac{\pi}{2} (k_+ + 1)$ for $k_+$ in the neighborhood of -1, in which case we put 
$q_+ = (k_+ + 1)$. This replacement
gives an expression that we call 
$\partial_+^2 \tilde{A}_{G,\Lambda} (\pi T, 1,0)$~:
\begin{multline}
\partial_+^2 \tilde{A}_{G,\Lambda} (\pi T, 1, 0) = 
\int d^3x \ x_+^2  \int d^3k_1 \ \frac{u_{\Lambda} (q_{1,+}) e^{i k_1 . x}}{i k_{1, 0} + \pi q_{1, +}  \cos \frac{\pi}{2} k_{1, -}}  \\
\int d^3k_2 \ \frac{u_{\Lambda} (q_{2,+}) e^{i k_2 . x}}{- i k_{2, 0} -
\pi q_{2, +} \cos \frac{\pi}{2} k_{2, -}}   \int d^3 k_3 \
\frac{u_{\Lambda} (q_{3,+}) e^{i k_3 . x}}{- i k_{3, 0} 
- \pi q_{3, +} 
\cos \frac{\pi}{2} k_{3, -}} e^{- i (\pi T x_0 + x_+)} \\
+ 2  \int d^3x \ x_+^2 \int d^3k_1 \ \frac{u_{\Lambda}(q_{1,+}) 
e^{i k_1 . x}}{i k_{1, 0} - \pi q_{1, +}  \cos \frac{\pi}{2}
k_{1, -}} \\
\int d^3k_2 \ \frac{u_{\Lambda} (q_{2,+}) e^{i k_2 . x}}{- i k_{2, 0} +
\pi q_{2, +}  \cos \frac{\pi}{2} k_{2, -}}  \int d^3k_3 \
\frac{u_{\Lambda} (q_{3,+}) e^{i k_3 . x}}{- i k_{3, 0} 
- \pi q_{3, +} \cos \frac{\pi}{2} k_{3, -}}  e^{- i (\pi T x_0 + x_+)} \ ,
\label{basequa}
\end{multline}
where the $u_{\Lambda} (q_{+})$'s is now the smooth
scaled cutoff function $u(M^{i_{max}(T) -\Lambda} q_+)$
which expresses the former sector constraint 
$s_+ \ge i_{max}(T) -\Lambda $ (recall that $u$ is our fixed
basic cutoff function).

In (\ref{basequa}) we can freely change each integral over 
$dk_+$ which ran over $[-2,2]$ into an integral on $dq_+$
which runs from $[-\infty,\infty]$. We still denote $\int d^3 k $
the corresponding integrals.

We write now for each propagator in (\ref{basequa}),
$u_{\Lambda} (q_{+}) = 1 + u^{1} (q_{+}) + u_{1}^{\Lambda} (q_{+})$ where
$u^{1} (q_{+}) =   u(q_+) -1 $ and $ u_{1}^{\Lambda} (q_{+}) = u_{\Lambda} (q_{+}) - u(q_+)$. In this way we generate three terms:

\begin{itemize}

\item one in which all three functions $u_{\Lambda} (q_{+})$  are replaced by 1.
We call this term $\partial_+^2 \tilde{A}_G (\pi T, 1, 0)$

\item one in which there is at least one factor $ u_{1}^{\Lambda} (q_{+})$
and no factor $u^{1} (q_{+})$. We call this term  
$\partial_+^2 A_{G, 1}^{ \Lambda} (\pi T, 1, 0)$.

\item finally one in which there is at least one factor $u^{1} (q_{+})$.
We call this term  
$\partial_+^2 A_{G}^1 (\pi T, 1, 0)$.
\end{itemize}

At this stage, we recapitulate~:
\begin{eqnarray}
\partial_+^2 A_G (\pi T, 1, 0) &=& 
\partial_+^2 \tilde{A}_G (\pi T, 1, 0)
+ \partial_+^2 A_{G, 1}^{\Lambda} (\pi T, 1, 0) +
\partial_+^2 A_{G}^1 (\pi T, 1, 0) 
\nonumber\\
&+& \Big( \partial_+^2 A_{G, \Lambda} (\pi T, 1, 0) - \partial_+^2 \tilde{A}_{G,\Lambda}  (\pi T, 1, 0) \Big)
+ \partial_+^2 A_G^\Lambda (\pi T, 1, 0) \ .
\end{eqnarray}
This relation shows that the quantity under study, $\partial_+^2 A_G (\pi T,1, 0)$, is equal to the approximation $\partial_+^2 \tilde{A}_G (\pi T, 1,0)$, up to the four error terms 
\begin{equation}
\partial_+^2 A_G^\Lambda (\pi T, 1, 0), \
\partial_+^2 A_{G, 1}^{\Lambda} (\pi T, 1, 0),\ 
\Big(\partial_+^2 A_{G,\Lambda} (\pi T, 1, 0) - \partial_+^2\tilde{A}_{G,\Lambda} (\pi T, 1 , 0) \Big),\ 
\partial_+^2 A_{G}^1 (\pi T, 1, 0)\ .
\end{equation}

Now we are going to prove a lower
bound similar to the one of Theorem \ref{main_theorem}, but on the quantity
$\partial_+^2 \tilde{A}_G (\pi T, 1, 0)$, and establish an upper bound on each of the four error terms. More precisely, if we have $\left| \partial_+^2 \tilde{A}_G (\pi T, 1, 0) \right| > \frac{K}{T}$ for 
some constant $K > 0$ and if the
modulus of each error term is smaller than $\frac{K'}{T}$ with $K' << K$, we shall conclude that~:
\begin{equation}
\left| \partial_+^2 A_G (\pi T, 1, 0) \right| > \frac{K - 4 K'}{T} \ ,
\end{equation}
which shall prove Theorem \ref{main_theorem}. The result that
$\left|\partial_+^2 \tilde{A}_G (\pi T, 1, 0) \right| > \frac{K}{T}$ is
really the most difficult to establish, and its proof is the heart of this
paper. But the control of the error terms is easier, and each one
will correspond to a lemma. We shall begin by these lemmas in next section,
and then turn to the lower bound on $\left| \partial_+^2 \tilde{A}_G (\pi T, 1, 0)\right|$.

\section{The control of the error terms}

First we state a result that is not
necessary for proving Theorem \ref{main_theorem} but whose proof illustrates
the way the sector decomposition allows us to establish quite easily upper bounds.

\begin{lemma}
There exists some constant $K_1 > 0$ such that~:
\begin{equation}
\left| \partial_+^2 A_G (\pi T, 1, 0) \right| \leq \frac{K_1}{T} \ .
\end{equation}
\label{1/T bound}
\end{lemma}

\medskip
\noindent \textbf{Proof~:}
\medskip
We use the decay property of $C_{(i, s_+, s_-)} (x)$ (see \cite{R1}, Lemma 1 )~:
\begin{equation}
\left| C_{(i, s_+, s_-)} (x) \right| \leq c. M^{-s_+ - s_-} \exp \left(- c
\Big(d_\sigma (x) \Big)^\alpha  \right)\ ,
\end{equation}
where $\alpha \in ]0 , 1 [$ is a fixed number, $c$ is a constant and $d_\sigma
(x) = M^{-i} |x_0| + M^{- s_+} |x_+| + M^{- s_-} |x_-|$.
We have~:
\begin{multline}
\left| \partial_+^2 A_G (k) \right| \leq c^3. M^{-\sum_{j=1}^3 s_j^+ - \sum_{j=1}^3 s_j^-}\\
\sum_{\{ i_j \}, \{ s_j^+ \}, \{ s_j^- \}} \int d^3x \ x_+^2 \exp 
\left(- c \sum_{j=1}^3 \Big( M^{-i_j} |x_0| + M^{s_j^+} |x_+| + M^{- s_-} 
|x_-| \Big)^\alpha \right) \ . 
\end{multline}

Among the indices $i_1$, $i_2$ and $i_3$, we keep the best one, i. e. the smallest one, to perform the integration over $x_0$. We proceed in an analogous way for the indices $(s_1^+, s_2^+, s_3^+)$ and $(s_1^-, s_2^-, s_3^-)$ respectively. Thus we have~:
\begin{equation}
\left| \partial_+^2 A_G (k) \right| \leq c^3   \sum_{\{ i_j \}, \{ s_j^+ \}, \{ s_j^- \}} M^{-\sum_{j=1}^3 s_j^+ - \sum_{j=1}^3 s_j^-} M^{\inf \{ i_j \}} M^{3 \inf \{  s_j^+ \}} M^{\inf \{ s_j^- \}} \ .
\label{formula}
\end{equation}
To carry out our discussion, we introduce several notations. If $(a_1, a_2, a_3)$ is a family of three (not necessarily distinct) real numbers, we denote as usual $\inf \{ a_j \}$ the smallest number among the $a_j$'s, but we define also 
\begin{equation}
\inf_2 \{ a_j \} = \inf \Big( \{ a_1, a_2, a_3 \} \backslash \{ \inf \{ a_1, a_2, a_3 \} \} \Big)
\end{equation}
and~:
\begin{equation}
\inf_3 \{ a_j \} = \inf \Big( \{ a_1, a_2, a_3 \} \backslash \{ \inf \{ a_1, a_2, a_3 \}, \inf_2 \{ a_1, a_2, a_3 \} \} \Big) \ .
\end{equation}
Remark that $\inf_3 \{ a_j \}$ is indeed $\sup \{ a_j \}$. Finally in this paragraph we shall write simply $\sum a_j$ instead of $\sum_{j = 1}^3 a_j$, and similarly for the $s_j^+$'s and  the $s_j^-$'s. With these notations, it is very easy to check the following identity~:
\begin{equation}
\inf \{ a_j \} = \frac{1}{3} \sum_{j=1}^3 a_j - \frac{1}{3} \Big[ \inf_2 \{
a_j \} - \inf \{ a_j \} \Big] - \frac{1}{3} \Big[ \inf_3 \{ a_j \} - \inf \{ a_j \} \Big] \ .
\end{equation}
We introduce the abbreviation~:
\begin{equation}
\Delta \{ a_j \} = \Big[ \inf_2 \{ a_j \} - \inf \{ a_j \} \Big] + \Big[ \inf_3 \{ a_j \} - \inf \{ a_j \} \Big] \ ,
\end{equation}
so that we have~:
\begin{equation}
\inf \{ a_j \} = \frac{1}{3} \sum a_j - \frac{1}{3} \Delta \{ a_j \} \ .
\label{important identity} 
\end{equation}
We use this identity to replace $\inf \{ i_j \}$ and $\inf \{ s_j^\pm \}$ in formula (\ref{formula}), and we obtain~:
\begin{equation}
\left| \partial_+^2 A_G (k) \right| \leq c^3 \sum_{\{ i_j \}, \{ s_j^+ \}, \{ s_j^- \}} M^{-\sum s_j^+ - \sum s_j^-} M^{\frac{1}{3} \sum i_j - \frac{1}{3} \Delta \{ i_j \}} M^{\sum s_j^+ - \Delta \{ s_j^+ \}} M^{\inf \{ s_j^- \}}\ .
\end{equation}
Since $\inf \{ s_j^- \} \leq \frac{1}{3} \sum s_j^-$, we can write~:
\begin{equation}
\left| \partial_+^2 A_G (k) \right| \leq c^3 \sum_{\{ i_j \}, \{ s_j^+ \}; \{ s_j^- \}} M^{\frac{1}{3} \sum i_j - \frac{1}{3} 
\Delta \{i_j\}}M^{-\Delta
\{ s_j^+ \}} M^{-\frac{2}{3} \sum s_j^-} \ .
\label{equation A}
\end{equation}
Now, we use the constraints in the sum $\sum_{\{ i_j \}, \{ s_j^+ \}; \{ s_j^- \}}$ to write, for each $j \in \{ 1, 2, 3 \}$~:
\begin{equation}
s_j^- \geq i_j - s_j^+ - 2 \ .
\end{equation}
We deduce that~:
\begin{equation}
\frac{1}{3} \sum s_j^- \geq \frac{1}{3} \sum i_j - \frac{1}{3} \sum s_j^+ - 2
\end{equation}
and
\begin{equation}
M^{-\frac{1}{3} \sum s_j^-} \leq M^2 M^{-\frac{1}{3} \sum i_j + \frac{1}{3} \sum s_j^+} \ .
\end{equation}
Replacing in equation (\ref{equation A}), we get~:
\begin{equation}
\left| \partial_+^2 A_G (k) \right| \leq c^3 M^2 \sum_{\{ i_j \}, \{ s_j^+ \}, \{ s_j^- \}} M^{-\frac{1}{3} \Delta \{ i_j \}} M^{\frac{1}{3} \sum s_j^+ - \Delta \{ s_j^+ \}} M^{-\frac{1}{3} \sum s_j^-} \ ,
\end{equation}
and using relation (\ref{important identity}), we have~:
\begin{equation}
\left| \partial_+^2 A_G (k) \right| \leq c^3 M^2 \sum_{\{ i_j \}, \{ s_j^+ \}, \{ s_j^- \}} M^{-\frac{1}{3} \Delta \{ i_j \}} M^{\inf \{ s_j^+ \} - \frac{2}{3} \Delta \{ s_j^+ \}} M^{-\frac{1}{3} \sum s_j^-} \ .
\label{lesinf}
\end{equation}
At last, let us denote $\kappa$ the value of the index $j$ such that $s_\kappa^+ = \inf \{ s_j^+ \}$. We write $\inf \{ s_j^+ \} = i_\kappa - (i_\kappa - s_\kappa^+)$. Finally we obtain~:
\begin{equation}
\left| \partial_+^2 A_G (k) \right| \leq c^3 M^2 \sum_{\{ i_j \}, \{ s_j^+ \}, \{ s_j^- \}} M^{i_\kappa} M^{-\frac{1}{3} \Delta \{ i_j \}} M^{- (i_\kappa - s_\kappa^+) - \frac{2}{3} \Delta\{ s_j^+ \}} M^{-\frac{1}{3} \sum s_j^-} \ .
\end{equation}

Clearly the sums over $s_1^-$, $s_2^-$ and $s_3^-$ can be bounded by $K_2 =
\frac{M}{(M^{1/3} - 1)^3}$. The decay $M^{-\frac{2}{3} \Delta \{ s_j^+ \}}$
can be used to perform the sums over $s_j^+$ for $j \neq \kappa$, also at a
cost $K_2$.
In the same way, we use the decay $M^{-\frac{1}{3} \Delta \{ i_j \}}$ to sum over the values $i_j$, $j \neq \kappa$ also at cost $K_2$ per sum.
It remains to sum over $s_\kappa^+$~:
\begin{equation}
\sum_{0 \leq s_\kappa^+ \leq i_\kappa} M^{- (i_\kappa - s_\kappa^+)} \leq \frac{M}{M - 1} \ .
\end{equation}
At last, we have~:
\begin{equation}
\left| \partial_+^2 A_G (k) \right|  \leq K \sum_{i_\kappa = 0}^{i_{max} (T)} M^{i_\kappa} =
K  \frac{M^{i_{max} (T) + 1}}{M - 1}
\end{equation}
and we have $M^{i_{max} (T)} \sim \frac{1}{T}$ (see \cite{R1}), which proves lemma \ref{1/T bound}.
\qed
\medskip

We have then the following lemma, which is a slight refinement of lemma \ref{1/T bound}~:
\begin{lemma}
\begin{equation}
\left| \partial_+^2 A_G^\Lambda (\pi T, 1, 0) \right| ,  
\left| \partial_+^2 A_{G, 1}^{\Lambda} (\pi T, 1, 0) \right|
\leq \frac{K_1}{M^\Lambda T}
\label{small 1/T bound}
\end{equation}
where $K_1$ is the constant of Lemma \ref{1/T bound}.
\end{lemma}

\medskip\noindent
\textbf{Proof~:}
It is similar to the proof of Lemma \ref{1/T bound}.
The case of $\partial_+^2 A_{G, 1}^{\Lambda} (\pi T, 1, 0)$
can be decomposed into sectors exactly
in the same way than $\partial_+^2 A_G^\Lambda (\pi T, 1, 0)$
because away from the singularity and in a bounded domain
in $k_+$, the presence of $\pi q_+$  instead
of  $\cos \frac{\pi}{2} k_+$ does not change anything to the
bounds on the propagators in sectors. 
Each step is then similar to the proof of 
of lemma \ref{1/T bound} until we arrive at the last sum, for which~:
\begin{eqnarray}
\sum_{i_\kappa = 0}^{i_{max} (T) - \Lambda} M^{i_\kappa} & = & \frac{M^{i_{max} (T) - \Lambda + 1} - 1}{M - 1} \\
& \leq & \frac{M}{M - 1} . \frac{M^{i_{max} (T)}}{M^\Lambda} =  
 \frac{K'}{T . M^\Lambda}\ ,
\end{eqnarray}
which proves the lemma. \qed

The following lemma bounds the contributions with at least
one large infrared cutoff $u^1$ on one propagator~:
\begin{lemma}
\begin{equation}
\left| \partial_+^2 A_G^1 (\pi T, 1, 0) \right| 
\leq  K_2
\label{highbound}
\end{equation}
where $K_2$ is some new constant.
\end{lemma}

\medskip\noindent
\textbf{Proof~:}
The main idea is that a propagator bearing cutoff $u^1 = 1- u$ on $q_+$
decays on a length scale $O(1)$ in $x_+$, so the factor
$x_+^2$ in $\partial_+^2 A_G^1$ is now harmless, and this prevents
the divergence in $1/T$ of the bound.

We remark first that in the amplitude $\partial_+^2 A_G^1$
we can change the sum over $x_+$ into a sum 
over the non zero values of $x_+$, because of the $x_+^2$ integrand.
Since a propagator bearing cutoff $u^1 = 1- u$ on $q_+$ is not absolutely integrable at large $q_+$, we first prepare 
all such propagators (there are between 1 and 3 of them)
using integration by parts.

For any such propagator we first split the $q_+$ integration into the
two regions $\int_{1}^{\infty} dq_+$ and $\int_{-\infty}^{-1}dq_+$
and treat only the first term, the other one being identical. Similarly
we can assume that we work on a 'right' propagator, so that
$q_+ = k_+ -1$, the other case being identical. The corresponding 
object is then: 
\begin{eqnarray}  D(x) &=&   e^{ix_+} 
\int dk_0 \int_{-2}^{2} dk_- \int_{1}^{\infty} d q_+
\ \frac{[1- u (q_{+})]  e^{i (k_0 x_0 + k_-x_- + q_+ x_+)}}
{i k_{0} + \pi q_{+}  \cos \frac{\pi}{2} k_{-}} 
\nonumber\\
&=& -\frac{i e^{ix_+}}{x_+} 
\int dk_0 \int_{-2}^{2} dk_- \int_{1}^{\infty} d q_+
\biggl[ \frac{[\pi \cos \frac{\pi}{2} k_{-}]
[1- u (q_{+}) ] e^{i (k_0 x_0 + k_-x_- + q_+ x_+)}}
{[i k_{0} + \pi q_{+}\cos \frac{\pi}{2} k_{-}]^2} 
\nonumber\\
&& \hskip7cm + \frac{ u' (q_{+})  e^{i (k_0 x_0 + k_-x_- + q_+ x_+)}}
{i k_{0} + \pi q_{+}\cos \frac{\pi}{2} k_{-}} \biggr] \ .
\end{eqnarray}

The last term, having a compact support 
$u'$ is similar to the ones of the  previous lemma,
and left to the reader. Let us treat the first term.

We divide it with a partition of unity
into new sectors $i,s_+, s_-$ according to the size of 
the denominator $i k_{0} + \pi q_{+}\cos \frac{\pi}{2} k_{-}$,
which is $M^{-i}$, the size of $q_+$ 
which is now of order $M^{+ s_+}$, with $s_+ >0$,
and of $k_-$ which is
of order $M^{-s_-} = M^{-i-s_+}$, with $s_- = i+s_+$.
The bounds are:
\begin{eqnarray} | D_{i,s_+, s_-}(x)| 
&\le & K | x_+ |^{-1}  M^{+i}  M^{s_+} M^{-2s_-}  
e^{-c[M^{-i}x_0 + M^{s_+}x_+ + M^{-s_-} x_-]^{\alpha} } 
\nonumber\\
&\le &  \frac{2K}{\pi} M^{-i-s_+} 
e^{-c[M^{-i}x_0 + M^{s_+}x_+ + M^{-s_-}x_-]^{\alpha}} \ ,
\end{eqnarray}
since for non zero $x_+$, on the tilted lattice $| x_+ |^{-1}$
is bounded by $2/\pi$. Hence taking into account that the
"integral"$\int dx_+$ is really a discrete sum on $\frac{\pi}{2} \mathbb{Z}$:
\begin{equation}
\int dx_+  x_+^2 | D_{i,s_+, s_-}(x)| \le  K M^{-i-3s_+} e^{-[M^{-i}x_0 + M^{s_+}x_+ + M^{-s_-}x_-]^{\alpha}/2} \ .
\end{equation}
Finally we need to optimize the $dx_0$ and $dx_-$ using the best 
of the three other propagators. This leads to a bound which 
obviously is uniform in $T$. For instance if the three
propagators have large infrared 
cutoffs $u^1 = 1- u$, we get the bound
\begin{equation}
\sum_{i_1,i_2,i_3 \atop s_{+,1},s_{+,2},s_{+,3}}  
K M^{- \sum_j i_j - \sum_j s_{+,j} -2 \sup s_{+,j} + 
\inf \{i \} + \inf \{i + s_+\}}
\le \sum_{i_1,i_2,i_3 \atop { s_{+,1} ,s_{+,2},s_{+,3}}} 
K M^{- (1/3)\sum_j i_j - (4/3)\sum_j s_{+,j} } \le K'\ ,
\end{equation}
and the other cases, when one or two propagators
are of ordinary type, are similar and left to the reader.
\qed

\medskip

Finally we state the lemma that allows us to control the replacement of $\cos \frac{\pi}{2} k_-$ by its Taylor expansion~:

\begin{lemma}
There exists a constant $K_3 > 0$ such that~:
\begin{equation}
\left| \partial_+^2 A_{G,\Lambda} (\pi T , 1, 0) - 
\partial_+^2 \tilde{A}_{G,\Lambda}
  (\pi T , 1 , 0) \right| \leq K_3 \ .
\end{equation}
\end{lemma}

\medskip\noindent
\textbf{Proof~:}
\

\begin{multline}
\partial_+^2 A_{G,\Lambda} (\pi T , 1 , 0) - 
\partial_+^2 \tilde{A}_{G,\Lambda} (\pi T , 1
, 0) \\
= \sum_{\{ \sigma_j \}, \ i_j, s_j^+ > i_{max} (T) - \Lambda \atop \sigma_1 \
\text{right}} \int d^3x \ x_+^2 \left[ C_{\sigma_1} (x) 
\bar{C}_{\sigma_2} (x) \bar{C}_{\sigma_3} (x) - \tilde{C}_{\sigma_1^r} (x)
\bar{\tilde{C}}_{\sigma_2^\ell} \bar{\tilde{C}}_{\sigma_3^\ell} (x) 
\right]
e^{- (\pi T x_0 + x_+)} \\
+ 2 \sum_{\{\sigma_j  \}, \ i_j, s_j^+ > i_{max} (T) - \Lambda \atop \sigma_2
\ \text{right}} \int d^3x \ x_+^2 \left[ C_{\sigma_1} (x) \bar{C}_{\sigma_2}
(x) \bar{C}_{\sigma_3} (x) - \tilde{C}_{\sigma_1^\ell} (x)
\bar{\tilde{C}}_{\sigma_2^r} \bar{\tilde{C}}_{\sigma_3^r} (x) \right]\ ,
\end{multline}
where
\begin{eqnarray}
\tilde{C}_{\sigma^r} (x) & = & \int d^3k \ \frac{u_{\sigma^r} (k) e^{i k . x}}{i k_0 + \pi (k_+
- 1) \cos \frac{\pi}{2} k_-} \\
\tilde{C}_{\sigma^\ell} (x) & = & \int d^3k \ \frac{u_{\sigma^l} (k) e^{i k . x}}{i k_0 - \pi
(k_+ + 1) \cos \frac{\pi}{2} k_-} \ .
\end{eqnarray}

Observing that there exists a constant $K_4$ such that~:
\begin{eqnarray}
\left| \cos \frac{\pi}{2} k_+ + \pi (k_+ - 1)  \right| & \leq & K_4 (k_+ -
1)^2 \\
\left| \cos \frac{\pi}{2} k_+ - \pi (k_+ + 1) \right| & \leq & K_4 (k_+ + 1)^2
\end{eqnarray}
uniformly in $k_+$, we have~:
\begin{equation}
\left| C_{\sigma^{r (\ell)}} (x) - \tilde{C}_{\sigma^{r (\ell)}} (x) \right|
\leq c' . M^{- 3 s_+ - s_-} e^{- c' d_\sigma^\alpha (x)} \ .
\end{equation}
Using the relation
\begin{equation}
C_{\sigma_1} \overline{C}_{\sigma_2} \overline{C}_{\sigma_3} -
\tilde{C}_{\sigma_1} \overline{\tilde{C}}_{\sigma_2}
\overline{\tilde{C}}_{\sigma_3} = (C_{\sigma_1} - \tilde{C}_{\sigma_1})
\overline{C}_{\sigma_2} \overline{C}_{\sigma_3} + \tilde{C}_{\sigma_1}
(\overline{C}_{\sigma_2} - \overline{\tilde{C}}_{\sigma_2}) C_{\sigma_3} +
\tilde{C}_{\sigma_1} \overline{\tilde{C}}_{\sigma_2} ( \overline{C}_{\sigma_3}
- \overline{\tilde{C}}_{\sigma_3}) \ ,
\end{equation}
to create differences of the type 
$C - \tilde{C}$, we gain $M^{- 2 s_+} \leq
M^{- 2 (i_{max} - \Lambda)}$ (provided $i_{max} (T) - 2 \Lambda \geq 0$, which we assume from now on) in the power counting with respect to a single
propagator. \qed

\medskip

At last, we state our main lower bound:

\begin{theorem}
\label{thecau}
There exists a constant $K_5 > 0$ such that~:
\begin{equation}
\left| \partial_+^2 \tilde{A}_G (\pi T , 1 , 0) \right| \geq
\frac{K_5}{T} \ .
\end{equation}
\end{theorem}

\medskip
This theorem with the lemmas of this section obviously imply Theorem
\ref{main_theorem}, hence the remaining of this paper is devoted to the proof of this Theorem \ref{thecau}.

\section{Integration over $k_{1,+}$, $k_{2,+}$ and $k_{3,+}$}
\subsection{Approximate expression}

We return to equation (\ref{basequa}),
in which all three cutoffs $u_{\Lambda}$ have been replaced by 1.
Let us write in equation (\ref{basequa}) 
$\partial_+^2 \tilde{A}_G (\pi T , 1 , 0)$ as 
$\partial_+^2 \tilde{A}_{G,1}  + 2 \partial_+^2 \tilde{A}_{G,2} $
and let us consider the first term $\partial_+^2 \tilde{A}_{G,1}$.

The first propagator (after a change of variable to call 
the dummy variable $q_+$ again $k_+$):
\begin{equation}
\int d^3k_1  \ \frac{e^{i k_1 . x} e^{i x_+}} {ik_{1,0}+\pi k_{1,+}\cos \left(\frac{\pi}{2}k_{1,-} \right)} \ .
\end{equation}  
For $\cos \left(\frac{\pi}{2}k_{1,-}\right) \neq 0$ we have~:
\begin{equation}
\int_{-\infty}^{+\infty}dk_{1,+} \ \frac{e^{ik_{1,+}x_+}}{ik_{1,0}+\pi k_{1,+} \cos \left(\frac{\pi}{2}k_{1,-}\right)} = \frac{1}{\pi \cos \left(\frac{\pi}{2}k_{1,-}\right)} \int_{-\infty}^{+\infty}dk_{1,+} \ \frac{e^{i k_{1,+}x_+}}{k_{1,+}+\left(\frac{i k_{1,0}}{\pi \cos (\frac{\pi}{2}k_{1,-})}\right)} \ . 
\end{equation}
The corresponding residue is $\exp \left({\frac{k_{1,0}x_+}{\pi \cos
\left(\frac{\pi}{2}k_{1,-}\right)}}\right)$. If $x_+ \ > 0$, then we 
move the path
of integration upwards. It is oriented in the positive direction,
so we get~: 
\begin{equation}
\chi (x_+ > 0) \, \chi \left (-\frac{k_{1,0}}{\pi \cos \left(\frac{\pi}{2}k_{1,-}\right)} > 0 \right ) 2i\pi \exp \left({\frac{k_{1,0}x_+}{\pi \cos
(\frac{\pi}{2}k_{1,-})}}\right) \ .
\end{equation}
If $x_+ < 0$, then the path of integration is moved downwards, and we
get a minus sign owing to the negative direction. Hence:
\begin{multline}
\int_{-\infty}^{+\infty}dk_{1,+} \ \frac{e^{ik_{1,+}x_+}}{ik_{1,0}+\pi
k_{1,+}\cos \left(\frac{\pi}{2}k_{1,-}\right)} = 
\frac{2i}{\cos
\left(\frac{\pi}{2}k_{1,-}\right)}\exp \left({\frac{k_{1,0}x_+}{\pi \cos
\left(\frac{\pi}{2}k_{1,-}\right)}}\right)  \\
\left [ \chi (x_+ > 0) \, \chi \left (-\frac{k_{1,0}}{\pi \cos
\left(\frac{\pi}{2}k_{1,-}\right)} > 0 \right ) - \chi (x_+ < 0) \, \chi 
\  \left (-\frac{k_{1,0}}{\pi \cos
\left(\frac{\pi}{2}k_{1,-}\right)} < 0 \right ) \right ] \ .
\end{multline}
We treat analogously the integrations over $k_{2,+}$ and
$k_{3,+}$. The only difference with the previous case is that these
propagators were near the left singularity $k_+ \simeq -1$, so there
are some sign changes in $q_{2,+}$ and $q_{3,+} \approx -1$. We obtain~:
\begin{multline}
\int_{-\infty}^{+\infty}dk_{2,+} \ \frac{e^{ik_{2,+}x_+}}{-ik_{2,0}-\pi
k_{2,+}\cos \left(\frac{\pi}{2}k_{2,-}\right)} = 
\frac{-2i}{\cos\left(\frac{\pi}{2}k_{2,-}\right)}\exp \left({-\frac{k_{2,0}x_+}{\pi \cos
(\frac{\pi}{2}k_{2,-})}}\right) \\
\left [ \chi (x_+ > 0) \, \chi \left (\frac{k_{2,0}}{\pi \cos
\left(\frac{\pi}{2}k_{2,-}\right)} < 0 \right ) - 
\chi (x_+ < 0) \, \chi \left (\frac{k_{2,0}}{\pi \cos
\left(\frac{\pi}{2}k_{2,-}\right)} > 0 \right ) \right ] \ .
\end{multline}
\begin{multline}
\partial_+^2 \tilde{A}_{G,1}(\pi T,1,0) = -8i\int d^3x
\int dk_{1,0} \,dk_{1,-} \,dk_{2,0} \,dk_{2,-} \,dk_{3,0} \,dk_{3,-} \\
x_+^2 \,\frac{\exp \left({\left ( \frac{k_{1,0}}{\pi \cos
\left(\frac{\pi}{2}k_{1,-}\right)}+\frac{k_{2,0}}{\pi \cos
\left(\frac{\pi}{2}k_{2,-}\right)}+\frac{k_{3,0}}{\pi \cos
\left(\frac{\pi}{2}k_{3,-}\right)} \right )x_+}\right)}{\cos
\left(\frac{\pi}{2}k_{1,-}\right)\cos \left(\frac{\pi}{2}k_{2,-}\right)\cos
(\frac{\pi}{2}k_{3,-})} \,e^{i(k_{1,0}+k_{2,0}+k_{3,0}+\pi T) 
x_0 }\,e^{i(k_{1,-}+k_{2,-}+k_{3,-})x_-}  \\
\left [ \chi (x_+ > 0) \,\chi \left (
\frac{k_{1,0}}{\pi \cos \left(\frac{\pi}{2}k_{1,-}\right)} < 0 \right ) \,\chi\left (\frac{k_{2,0}}{\pi \cos \left(\frac{\pi}{2}k_{2,-}\right)} 
< 0 \right )\chi \left(\frac{k_{3,0}}{\pi \cos \left(\frac{\pi}{2}k_{3,-}\right)} < 0 \right ) \right. \\
- \left.  \chi (x_+ < 0) \,\chi \left (\frac{k_{1,0}}{\pi \cos
\left(\frac{\pi}{2}k_{1,-}\right)} > 0 \right ) \,\chi \left
(\frac{k_{2,0}}{\pi \cos \left(\frac{\pi}{2}k_{2,-}\right)} > 0 
\right )\chi
\left (\frac{k_{3,0}}{\pi \cos \left(\frac{\pi}{2}k_{3,-}\right)} > 0 
\right ) \right ]\ .
\end{multline}

\subsection{Integration over $x_0$ and $k_{3,0}$}

The calculation is done integrating over $x_0$, which leads to a
delta function in the integrand, denoted with a slight abuse of
notation by $\delta (k_{1,0}+k_{2,0}+k_{3,0}+ \pi T=0)$. In fact,
there is a prefactor $\frac{1}{T}$ that compensates the $T$ factor of
$\int dk_{3,0}$~: remember that $\int dk_{3,0}$ means precisely~:
$2\pi T \sum_{k_{3,0} \in \pi T + 2\pi T \mathbb{Z}}$. This yields~:
\begin{multline}
\partial_+^2 \tilde{A}_{G,1}(\pi T,1,0) = -8i\int dx_+ \,dx_-
\int dk_{1,0} \,dk_{1,-} \,dk_{2,0} \,dk_{2,-} \,dk_{3,0} \,dk_{3,-} \\
x_+^2 \,\frac{e^{\left ( \frac{k_{1,0}}{\pi \cos
(\frac{\pi}{2}k_{1,-})}+\frac{k_{2,0}}{\pi \cos
(\frac{\pi}{2}k_{2,-})}+\frac{k_{3,0}}{\pi \cos
(\frac{\pi}{2}k_{3,-})} \right )x_+}}{\cos
(\frac{\pi}{2}k_{1,-})\cos (\frac{\pi}{2}k_{2,-})\cos
(\frac{\pi}{2}k_{3,-})}\,e^{i(k_{1,-}+k_{2,-}+k_{3,-})x_-}\,\delta 
(k_{1,0}+k_{2,0}+k_{3,0}+ \pi T=0) \\
\left [ \chi (x_+ > 0) \,\chi (
\frac{k_{1,0}}{\pi \cos (\frac{\pi}{2}k_{1,-})} < 0) \,\chi (
\frac{k_{2,0}}{\pi \cos (\frac{\pi}{2}k_{2,-})} < 0)\chi (
\frac{k_{3,0}}{\pi \cos (\frac{\pi}{2}k_{3,-})} < 0) \right. \\
 - \left.  \chi (x_+ < 0) \,\chi (\frac{k_{1,0}}{\pi \cos 
(\frac{\pi}{2}k_{1,-})} > 0) \,\chi (\frac{k_{2,0}}{\pi \cos 
(\frac{\pi}{2}k_{2,-})} > 0)\chi (\frac{k_{3,0}}{\pi \cos 
(\frac{\pi}{2}k_{3,-})} > 0) \right ]\ .
\end{multline}

At this stage, we can use the delta function to integrate, for
instance, over $k_{3,0}$~:
\begin{multline}
\partial_+^2 \tilde{A}_{G,1}(\pi T,1,0) = -8i\int dx_+ \,dx_-
\int dk_{1,0} \,dk_{1,-} \,dk_{2,0} \,dk_{2,-} \,dk_{3,-} \\
x_+^2 \,\frac{e^{\left ( \frac{k_{1,0}}{\pi \cos
(\frac{\pi}{2}k_{1,-})}+\frac{k_{2,0}}{\pi \cos
(\frac{\pi}{2}k_{2,-})}-\frac{k_{1,0}+k_{2,0}+\pi T}{\pi \cos
(\frac{\pi}{2}k_{3,-})} \right )x_+}}{\cos
(\frac{\pi}{2}k_{1,-})\cos (\frac{\pi}{2}k_{2,-})\cos
(\frac{\pi}{2}k_{3,-})}\,e^{i(k_{1,-}+k_{2,-}+k_{3,-})x_-} \\
\left [ \chi (x_+ > 0) \,\chi \left (
\frac{k_{1,0}}{\pi \cos (\frac{\pi}{2}k_{1,-})} < 0 \right ) \,\chi
\left (
\frac{k_{2,0}}{\pi \cos (\frac{\pi}{2}k_{2,-})} < 0 \right ) \,\chi
\left (
\frac{k_{1,0}+k_{2,0}+\pi T}{\pi \cos (\frac{\pi}{2}k_{3,-})} > 0
\right ) \right. \\
 - \left.  \chi (x_+ < 0) \,\chi \left (\frac{k_{1,0}}{\pi \cos
(\frac{\pi}{2}k_{1,-})} > 0 \right ) \,\chi \left
(\frac{k_{2,0}}{\pi \cos (\frac{\pi}{2}k_{2,-})} > 0 \right )
\,\chi \left (\frac{k_{1,0}+k_{2,0}+\pi T}{\pi \cos
(\frac{\pi}{2}k_{3,-})} < 0 \right ) \right ]\ .
\end{multline}

\subsection{Simplification}

This rather complicated expression can be slightly simplified. Indeed,
if we perform the change of variables~:
\begin{equation}
\left\{
\begin{array}{ccc}
x_+' & = & -x_+ \\
k_{1,0}' & = & -k_{1,0} \\
k_{2,0}' & = & -k_{2,0}
\end{array}
\right.
\end{equation}
the integral
\begin{multline}
\int dx_+ \,dx_-
\int dk_{1,0} \,dk_{1,-} \,dk_{2,0} \,dk_{2,-} \,dk_{3,-} \,\,
x_+^2 \,\frac{e^{\left ( \frac{k_{1,0}}{\pi \cos
(\frac{\pi}{2}k_{1,-})}+\frac{k_{2,0}}{\pi \cos
(\frac{\pi}{2}k_{2,-})}-\frac{k_{1,0}+k_{2,0}+\pi T}{\pi \cos
(\frac{\pi}{2}k_{3,-})} \right )x_+}}{\cos
(\frac{\pi}{2}k_{1,-})\cos (\frac{\pi}{2}k_{2,-})\cos
(\frac{\pi}{2}k_{3,-})} \\
\chi (x_+ < 0) \,\chi \left (\frac{k_{1,0}}{\pi \cos
(\frac{\pi}{2}k_{1,-})} > 0 \right ) \,\chi \left
(\frac{k_{2,0}}{\pi \cos (\frac{\pi}{2}k_{2,-})} > 0 \right )
\,\chi \left (\frac{k_{1,0}+k_{2,0}+\pi T}{\pi \cos
(\frac{\pi}{2}k_{3,-})} < 0 \right )
\end{multline}
becomes~:
\begin{multline}
\int dx_+' \,dx_-
\int dk_{1,0}' \,dk_{1,-} \,dk_{2,0}' \,dk_{2,-} \,dk_{3,-} \,\,
x_+'^2 \,\frac{e^{\left ( \frac{k_{1,0}'}{\pi \cos
    (\frac{\pi}{2}k_{1,-})}+\frac{k_{2,0}'}{\pi \cos
    (\frac{\pi}{2}k_{2,-})}-\frac{k_{1,0}'+k_{2,0}'-\pi T}{\pi \cos
    (\frac{\pi}{2}k_{3,-})} \right )x_+}}{\cos
  (\frac{\pi}{2}k_{1,-})\cos (\frac{\pi}{2}k_{2,-})\cos
  (\frac{\pi}{2}k_{3,-})} \\
\chi (x_+' > 0) \,\chi \left (\frac{k_{1,0}'}{\pi \cos
       (\frac{\pi}{2}k_{1,-})} < 0 \right ) \,\chi \left
     (\frac{k_{2,0}'}{\pi \cos (\frac{\pi}{2}k_{2,-})} < 0 \right )
   \,\chi \left (\frac{k_{1,0}'+k_{2,0}'-\pi T}{\pi \cos
       (\frac{\pi}{2}k_{3,-})} > 0 \right ) \ .
\end{multline}
Consequently the previous expression of $\partial_+^2
\tilde{A}_{G,1}(\pi T,1,0)$ can be factorized~:
\begin{multline}
\partial_+^2 \tilde{A}_{G,1}(\pi T,1,0) = -8i\int dx_+ \,dx_-
\int dk_{1,0} \,dk_{1,-} \,dk_{2,0} \,dk_{2,-} \,dk_{3,-} \\
x_+^2 \,\frac{e^{\left ( \frac{k_{1,0}}{\pi \cos
(\frac{\pi}{2}k_{1,-})}+\frac{k_{2,0}}{\pi \cos
(\frac{\pi}{2}k_{2,-})}-\frac{k_{1,0}+k_{2,0}}{\pi \cos
(\frac{\pi}{2}k_{3,-})} \right )x_+}}{\cos
(\frac{\pi}{2}k_{1,-})\cos (\frac{\pi}{2}k_{2,-})\cos
(\frac{\pi}{2}k_{3,-})}\,e^{i(k_{1,-}+k_{2,-}+k_{3,-})x_-} \\
\chi (x_+ > 0) \,\chi \left (
\frac{k_{1,0}}{\pi \cos (\frac{\pi}{2}k_{1,-})} < 0 \right ) \,\chi
\left (
\frac{k_{2,0}}{\pi \cos (\frac{\pi}{2}k_{2,-})} < 0 \right ) \\ 
\left [ e^{\frac{-Tx_+}{\cos (\frac{\pi}{2}(k_{3,-}))}}\chi
\left (
\frac{k_{1,0}+k_{2,0}+\pi T}{\pi \cos (\frac{\pi}{2}k_{3,-})} > 0
\right ) - e^{\frac{Tx_+}{\cos (\frac{\pi}{2}(k_{3,-}))}}\chi \left 
(\frac{k_{1,0}+k_{2,0}-\pi T}{\pi \cos
(\frac{\pi}{2}k_{3,-})} > 0 \right ) \right ] \ .
\end{multline} 

\section{Integration over $x_-$ and $k_{3,-}$}

We now are going to perform the integration over $x_-$, which will provide a conservation rule for the moments $k_{1,0}$, $k_{2,0}$ and $k_{3,0}$,
but only modulo 2. To understand that, remember that $\int
dx_+ \, dx_-$ means more precisely~: 
$\sum'_{(x_+,x_-) \in \left( \frac{\pi}{2}\mathbb{Z} \right )^2 }$, where the prime in the sum means that one has to respect a parity condition 
between $x_+$ and $x_-$. By slight abuse of
language, we say that $x_+$ and $x_-$ have the same parity when $x_+ +
x_- \in \pi \mathbb{Z}$ . So 
$ \sum'_{ (x_+,x_-) \in \left( \frac{\pi}{2}\mathbb{Z} \right )^2 }
$does not mean~:
$\sum_{x_+ \in \frac{\pi}{2}\mathbb{Z}}\sum_{x_- \in
\frac{\pi}{2}\mathbb{Z}}$
but
$\sum_{x_+ \in \pi \mathbb{Z}}\sum_{x_- \in \pi \mathbb{Z}} + \sum_{x_+ \in
\frac{\pi}{2}+\pi \mathbb{Z}}\sum_{x_- \in \frac{\pi}{2} + \pi
\mathbb{Z}}$.
Now, 
\begin{equation}
\sum_{x_- \in \pi \mathbb{Z}}e^{i(k_{1,-}+k_{2,-}+k_{3,-})x_-} = \delta
(k_{1,-}+k_{2,-}+k_{3,-} = 0 [2])
\end{equation}  
where by $\delta
(k_{1,-}+k_{2,-}+k_{3,-} = 0 [2])$, we denote~: $\sum_{n \in \mathbb{Z}}\delta
(k_{1,-}+k_{2,-}+k_{3,-} = 2n)$.
Then it is clear that 
\begin{equation}
\sum_{x_- \in \frac{\pi}{2} + \pi \mathbb{Z}}e^{i(k_{1,-}+k_{2,-}+k_{3,-})x_-} = e^{i\frac{\pi}{2}(k_{1,-}+k_{2,-}+k_{3,-})}\,\delta
(k_{1,-}+k_{2,-}+k_{3,-} = 0 [2])\ .
\end{equation} 
Indeed, the factor
$e^{i\frac{\pi}{2}(k_{1,-}+k_{2,-}+k_{3,-})}$ can take only two
values~: $1$ if $k_{1,-}+k_{2,-}+k_{3,-} = 0[4]$, and $-1$ if
$k_{1,-}+k_{2,-}+k_{3,-} = 2[4]$. Hence it is convenient to
distinguish these two cases and write~:
\begin{equation}
\delta(k_{1,-}+k_{2,-}+k_{3,-} = 0 [2]) =
\delta(k_{1,-}+k_{2,-}+k_{3,-} = 0 [4]) +  \delta(k_{1,-}+k_{2,-}+k_{3,-} = 2 [4]) 
\end{equation} 
and
\begin{eqnarray}
&&e^{i\frac{\pi}{2}(k_{1,-}+k_{2,-}+k_{3,-})}\delta(k_{1,-}+k_{2,-}+k_{3,-} = 0 [2]) 
\nonumber\\
&&=
\delta(k_{1,-}+k_{2,-}+k_{3,-} = 0 [4]) -  \delta(k_{1,-}+k_{2,-}+k_{3,-} = 2 [4])\ .
\end{eqnarray} 
At this stage, we can gather the previous remarks in the following
formula :
\begin{multline}
\partial_+^2\tilde{A}_{G,1}(\pi T,1,0) = -8i\sum_{x_+ \in
\frac{\pi}{2}\mathbb{N}^*} \int dk_{1,0} \, dk_{2,0} \, dk_{1,-} \, 
dk_{2,-}\, dk_{3,-} \\
x_+^2 \,\frac{e^{\left ( \frac{k_{1,0}}{\pi \cos
    (\frac{\pi}{2}k_{1,-})}+\frac{k_{2,0}}{\pi \cos
    (\frac{\pi}{2}k_{2,-})}-\frac{k_{1,0}+k_{2,0}}{\pi \cos
    (\frac{\pi}{2}k_{3,-})} \right )x_+}}{\cos
  (\frac{\pi}{2}k_{1,-})\cos (\frac{\pi}{2}k_{2,-})\cos
  (\frac{\pi}{2}k_{3,-})}\,\delta(k_{1,-}+k_{2,-}+k_{3,-}=0[4]) \\ 
\chi \left (
\frac{k_{1,0}}{\pi \cos (\frac{\pi}{2}k_{1,-})} < 0 \right ) \,\chi
\left (
\frac{k_{2,0}}{\pi \cos (\frac{\pi}{2}k_{2,-})} < 0 \right ) \\
\left [ e^{\frac{-Tx_+}{\cos (\frac{\pi}{2}(k_{3,-}))}}\chi
  \left (
\frac{k_{1,0}+k_{2,0}+\pi T}{\pi \cos (\frac{\pi}{2}k_{3,-})} > 0
\right ) 
 - e^{\frac{Tx_+}{\cos (\frac{\pi}{2}(k_{3,-}))}}\chi \left 
(\frac{k_{1,0}+k_{2,0}-\pi T}{\pi \cos
       (\frac{\pi}{2}k_{3,-})} > 0 \right ) \right ] \\
-8i\sum_{x_+ \in
\frac{\pi}{2}\mathbb{N}^*} \int dk_{1,0} \, dk_{2,0} \, dk_{1,-} \, 
dk_{2,-} \, dk_{3,-} \\
x_+^2 \,\frac{e^{\left ( \frac{k_{1,0}}{\pi \cos
(\frac{\pi}{2}k_{1,-})}+\frac{k_{2,0}}{\pi \cos
(\frac{\pi}{2}k_{2,-})}-\frac{k_{1,0}+k_{2,0}}{\pi \cos
(\frac{\pi}{2}k_{3,-})} \right )x_+}}{\cos
(\frac{\pi}{2}k_{1,-})\cos (\frac{\pi}{2}k_{2,-})\cos
(\frac{\pi}{2}k_{3,-})}\,\delta(k_{1,-}+k_{2,-}+k_{3,-}=2[4]) \\ 
[\chi(x_+ \,\, even) - \chi(x_+ \,\, odd)]\chi \left (
\frac{k_{1,0}}{\pi \cos (\frac{\pi}{2}k_{1,-})} < 0 \right ) \,\chi
\left (
\frac{k_{2,0}}{\pi \cos (\frac{\pi}{2}k_{2,-})} < 0 \right ) \\
\left [ e^{\frac{-Tx_+}{\cos (\frac{\pi}{2}(k_{3,-}))}}\chi
\left (
\frac{k_{1,0}+k_{2,0}+\pi T}{\pi \cos (\frac{\pi}{2}k_{3,-})} > 0
\right )  - e^{\frac{Tx_+}{\cos (\frac{\pi}{2}(k_{3,-}))}}\chi \left 
(\frac{k_{1,0}+k_{2,0}-\pi T}{\pi \cos
(\frac{\pi}{2}k_{3,-})} > 0 \right ) \right ]\ .
\end{multline}
Then we can perform the integration over $k_{3,-}$. Formally, we only need to replace $\cos (\frac{\pi}{2}k_{3,-})$ by $\cos
(\frac{\pi}{2}(k_{1,-}+k_{2,-}))$ for the first piece and with $-\cos
(\frac{\pi}{2}(k_{1,-}+k_{2,-}))$ for the second piece. We obtain~:
\begin{multline}
\partial_+^2\tilde{A}_{G,1}(\pi T,1,0) = -8i\sum_{x_+ \in
\frac{\pi}{2}\mathbb{N}^*} \int dk_{1,0} \, dk_{2,0} \, dk_{1,-} \,  dk_{2,-}
\\
x_+^2 \,\frac{e^{\left ( \frac{k_{1,0}}{\pi \cos
(\frac{\pi}{2}k_{1,-})}+\frac{k_{2,0}}{\pi \cos
(\frac{\pi}{2}k_{2,-})}-\frac{k_{1,0}+k_{2,0}}{\pi \cos
(\frac{\pi}{2}(k_{1,-}+k_{2,-}))} \right )x_+}}{\cos
(\frac{\pi}{2}k_{1,-})\cos (\frac{\pi}{2}k_{2,-})\cos
(\frac{\pi}{2}(k_{1,-}+k_{2,-}))} 
\chi \left (
\frac{k_{1,0}}{\pi \cos (\frac{\pi}{2}k_{1,-})} < 0 \right ) \,\chi
\left (
\frac{k_{2,0}}{\pi \cos (\frac{\pi}{2}k_{2,-})} < 0 \right ) 
\\
\bigg[ e^{\frac{-Tx_+}{\cos (\frac{\pi}{2}(k_{1,-}+k_{2,-}))}}\chi
\left (
\frac{k_{1,0}+k_{2,0}+\pi T}{\pi \cos (\frac{\pi}{2}(k_{1,-}+k_{2,-}))} > 0
\right ) 
\\ 
- e^{\frac{Tx_+}{\cos (\frac{\pi}{2}(k_{1,-}+k_{2,-}))}}\chi \left 
(\frac{k_{1,0}+k_{2,0}-\pi T}{\pi \cos
(\frac{\pi}{2}(k_{1,-}+k_{2,-}))} > 0 \right ) \bigg] 
\\
+8i\sum_{x_+ \in
\frac{\pi}{2}\mathbb{N}^*} \int dk_{1,0} \, dk_{2,0} \, dk_{1,-} \, 
dk_{2,-}  
x_+^2 \,\frac{e^{\left ( \frac{k_{1,0}}{\pi \cos
(\frac{\pi}{2}k_{1,-})}+\frac{k_{2,0}}{\pi \cos
(\frac{\pi}{2}k_{2,-})}+\frac{k_{1,0}+k_{2,0}}{\pi \cos
(\frac{\pi}{2}(k_{1,-}+k_{2,-}))} \right )x_+}}{\cos
(\frac{\pi}{2}k_{1,-})\cos (\frac{\pi}{2}k_{2,-})\cos
(\frac{\pi}{2}(k_{1,-}+k_{2,-}))} 
\\ 
[\chi(x_+ \,\, even) - \chi(x_+ \,\, odd)]\chi \left (
\frac{k_{1,0}}{\pi \cos (\frac{\pi}{2}k_{1,-})} < 0 \right ) \,\chi
\left (
\frac{k_{2,0}}{\pi \cos (\frac{\pi}{2}k_{2,-})} < 0 \right ) 
\\
\bigg[ e^{\frac{Tx_+}{\cos (\frac{\pi}{2}(k_{1,-}+k_{2,-}))}}\chi
\left (
\frac{k_{1,0}+k_{2,0}+\pi T}{\pi \cos (\frac{\pi}{2}(k_{1,-}+k_{2,-}))} < 0
\right ) 
\\ 
- e^{\frac{-Tx_+}{\cos (\frac{\pi}{2}(k_{1,-}+k_{2,-}))}}\chi \left 
(\frac{k_{1,0}+k_{2,0}-\pi T}{\pi \cos
       (\frac{\pi}{2}(k_{1,-}+k_{2,-}))} < 0 \right ) \bigg]\ .
\end{multline}
Now it is clear that $\partial_+^2\tilde{A}_{G,1}(\pi T,1,0)$ is a
purely imaginary number. The first
piece gives the
leading behavior as $T \rightarrow 0$. Indeed the second piece is much
smaller, thanks to the compensation in
$[\chi(x_+ \,\, even) - \chi(x_+ \,\, odd)]$.
Indeed the sum 
\begin{equation}
\sum_{x_+ \in\frac{\pi}{2}\mathbb{N}^*}  x_+^2 [\chi(x_+ \,\, even) - \chi(x_+ \,\, odd)]...
\end{equation} 
can be written as a sum of two terms of the type
\begin{equation}
\sum_{n \in \mathbb{N}^*} \int dk e^{-2A(k)n}[(2n)^2- (2n+1)^2 e^{-A(k)}]B(k)
\end{equation}
where $A$ and $B$ are independent of $n$ and $A(k)>0$.
Then we can decompose the remaining integrals $\int dk$ into two zones,
according to whether $A(k) \ge T^{1/3}$ or $A(k) \le T^{1/3}$. In the first zone we do not need to exploit the subtraction, but we have simply
$\sum_{n \in \mathbb{N}^*}  n^2e^{-2T^{1/3}n} \le c.T^{-2/3} << T^{-1}$,
and in the second zone, we use $|(2n)^2- (2n+1)^2 e^{-A(k)}| \le 4n+1 + (2n+1)^2 A(k) \le 4n+1 + (2n+1)^2 T^{1/3} $. The first term in $4n+1$
is then bounded with the same techniques than Lemma \ref{1/T bound},
but the factor $M^{\inf \{i_j\}+ 3 \inf \{ s_j^+  \} + \inf\{s_j^- \} }$
is replaced by $M^{\inf \{i_j\}+ 2\inf \{ s_j^+  \} + \inf\{s_j^- \} }$
and the bound corresponding to equation (\ref{lesinf}) gives now
\begin{equation}
\sum_{\{ i_j \}, \{ s_j^+ \}, \{ s_j^- \}} M^{-\frac{1}{3} \Delta \{ i_j \}} M^{- \frac{2}{3} \Delta \{ s_j^+ \}} M^{-\frac{1}{3} \sum s_j^-} \le 0(1) \ ,
\end{equation}
hence this piece does not diverge at all when $T \to 0$. Finally
the piece with the factor $(2n+1)^2 T^{1/3} $ 
is similar to previous pieces, except for the
new factor $T^{1/3}$, so that it is bounded in the manner of Lemma \ref{1/T bound} by a factor $c.T^{-1}T^{1/3}=c.T^{-2/3}$.

So we are left to study~:
\begin{multline}
A_1(T) = -8i\sum_{x_+ \in
\frac{\pi}{2}\mathbb{N}^*} \int dk_{1,0} \, dk_{2,0} \, dk_{1,-} \, dk_{2,-} x_+^2 \,\frac{e^{\left ( \frac{k_{1,0}}{\pi \cos
(\frac{\pi}{2}k_{1,-})}+\frac{k_{2,0}}{\pi \cos
(\frac{\pi}{2}k_{2,-})}-\frac{k_{1,0}+k_{2,0}}{\pi \cos
(\frac{\pi}{2}(k_{1,-}+k_{2,-}))} \right )x_+}}{\cos
(\frac{\pi}{2}k_{1,-})\cos (\frac{\pi}{2}k_{2,-})\cos
(\frac{\pi}{2}(k_{1,-}+k_{2,-}))} 
\\ 
\chi \left (
\frac{k_{1,0}}{\pi \cos (\frac{\pi}{2}k_{1,-})} < 0 \right ) \,\chi
\left (\frac{k_{2,0}}{\pi \cos (\frac{\pi}{2}k_{2,-})} < 0 \right ) \\
\bigg[ e^{\frac{-Tx_+}{\cos (\frac{\pi}{2}(k_{1,-}+k_{2,-}))}}\chi
\left (
\frac{k_{1,0}+k_{2,0}+\pi T}{\pi \cos (\frac{\pi}{2}(k_{1,-}+k_{2,-}))} > 0
\right )
\\ 
- e^{\frac{Tx_+}{\cos (\frac{\pi}{2}(k_{1,-}+k_{2,-}))}}\chi 
\left(\frac{k_{1,0}+k_{2,0}-\pi T}{\pi \cos
(\frac{\pi}{2}(k_{1,-}+k_{2,-}))} > 0 \right ) \bigg] \ .
\end{multline}

\section{Leading contribution}
\subsection{Symmetry properties}

Henceforward, we shall denote the integrand by
$F(x_+,k_{1,0},k_{2,0},k_{1,-},k_{2,-})$ so that~:
\begin{equation}
A_1(T)=-8i\sum_{x_+ \in \frac{\pi}{2}\mathbb{N}^*} \int dk_{1,0} \, dk_{2,0} \,
dk_{1,-} \, dk_{2,-} \,F(x_+,k_{1,0},k_{2,0},k_{1,-},k_{2,-})\ .
\end{equation}
The couple of variables of integration $(k_{1,-},k_{2,-})$ describes
the square $[-2,2]^2$. To pursue the calculation, we shall make a
partition of $[-2,2]^2$, according to the signs of $\cos
(\frac{\pi}{2}k_{1,-})$, $\cos (\frac{\pi}{2}k_{2,-})$ and $\cos
(\frac{\pi}{2}(k_{1,-}+k_{2,-}))$. This partition is represented in
Figure \ref{domaine}:
\begin{figure}[H]
\centerline{\epsfig{figure=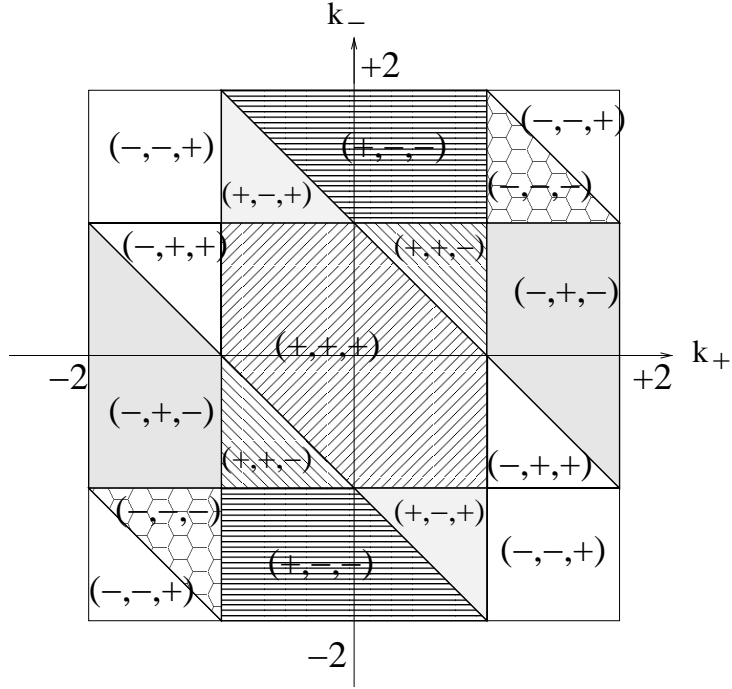,width=12cm,angle=270}}
\caption{The domain of integration in $(k_+ , k_-)$}
\label{domaine}
\end{figure}

The signs of the three cosines determine eight cases we can discuss
separately. In fact, it is possible to restrict the domain of
integration thanks to symmetries of the integrand involving the
variables $k_{1,-}$ and $k_{2,-}$ together with the variables
$k_{1,0}$ and $k_{2,0}$, which describe independently the set $\pi T +
2\pi T\mathbb{Z}$.

It is evident, by the parity of the cosine function, that the
integrand is invariant under the replacement $k_{1,-} \rightarrow
-k_{1,-}$ and $k_{2,-} \rightarrow -k_{2,-}$, which corresponds to the
central symmetry with respect to the origin (0,0). Hence we have~:
\begin{equation}
A_1(T) = -16i\sum_{x_+ \in
\frac{\pi}{2}\mathbb{N}^*} \int dk_{1,0} \, dk_{2,0} \int_{-2}^{2} 
dk_{1,-}
\int_{0}^{2} dk_{2,-} \,F(x_+,k_{1,0},k_{2,0},k_{1,-},k_{2,-})\ .
\end{equation}
Symmetry properties of $F(x_+,k_{1,0},k_{2,0},k_{1,-},k_{2,-})$ can be
exploited further. The above integral may be separated into two pieces~:
\begin{multline}
A_1(T) = -16i\left(\sum_{x_+ \in
\frac{\pi}{2}\mathbb{N}^*} \int dk_{1,0} \, dk_{2,0} \int_{-2}^{0} 
dk_{1,-}
\int_{0}^{2} dk_{2,-} \,F(x_+,k_{1,0},k_{2,0},k_{1,-},k_{2,-})
\right. \\
+ \left. \sum_{x_+ \in
\frac{\pi}{2}\mathbb{N}^*} \int dk_{1,0} \, dk_{2,0} \int_{0}^{2} dk_{1,-}
\int_{0}^{2} dk_{2,-} \,F(x_+,k_{1,0},k_{2,0},k_{1,-},k_{2,-}) \right )\ .
\end{multline}
For the first integral, one can easily verify that the integrand
$F(x_+,k_{1,0},k_{2,0},k_{1,-},k_{2,-})$ is invariant under the change
of variables~:
\begin{equation}
k_{1,0}' =  k_{2,0} \ , \
k_{2,0}' =  k_{1,0} \ , \
k_{1,-}' =  - k_{2,-} \ , \
k_{2,-}' =  - k_{1,-} \ .
\end{equation}
We get~:
\begin{multline}
\int dk_{1,0} \, dk_{2,0} \int_{-2}^{0} dk_{1,-}
  \int_{0}^{2} dk_{2,-} \,F(x_+,k_{1,0},k_{2,0},k_{1,-},k_{2,-}) = \\
2\int dk_{1,0} \, dk_{2,0} \int_{-2}^{0} dk_{1,-}
  \int_{-k_{1,-}}^{2} dk_{2,-} \,F(x_+,k_{1,0},k_{2,0},k_{1,-},k_{2,-})\ .
\end{multline}
We treat analogously the other integral~; we set~:
\begin{equation}
k_{1,0}'  =  k_{2,0} \ , \
k_{2,0}'  =  k_{1,0} \ , \
k_{1,-}'  =  k_{2,-} \ , \
k_{2,-}'  =  k_{1,-} \ .
\end{equation}
Hence~:
\begin{multline}
\int dk_{1,0} \, dk_{2,0} \int_{0}^{2} dk_{1,-}
  \int_{0}^{2} dk_{2,-} \,F(x_+,k_{1,0},k_{2,0},k_{1,-},k_{2,-}) = \\
2\int dk_{1,0} \, dk_{2,0} \int_{0}^{2} dk_{1,-}
  \int_{k_{1,-}}^{2} dk_{2,-} \,F(x_+,k_{1,0},k_{2,0},k_{1,-},k_{2,-})\ .
\end{multline}

Finally, we have established owing to symmetry properties that~:
\begin{equation}
A_1(T) = -32i\int dk_{1,0} \, dk_{2,0} \iint\limits_{\mathcal{T}} dk_{1,-} \, dk_{2,-} \,F(x_+,k_{1,0},k_{2,0},k_{1,-},k_{2,-}) \ ,
\end{equation}
the domain of integration in $(k_{1,-},k_{2,-})$ being the triangle $\mathcal{T}$
delimited by the lines $k_{2,-}=2$, $k_{2,-}=k_{1,-}$ and
$k_{2,-}=-k_{1,-}$.

\subsection{Discussion of the various cases}
\subsubsection{The $(+,+,+)$ case}

As we have said, it is now convenient to carry a discussion about the
signs of $\cos (\frac{\pi}{2}k_{1,-})$, $\cos (\frac{\pi}{2}k_{2,-})$
and $\cos (\frac{\pi}{2}(k_{1,-}+k_{2,-}))$, which allows us to
perform explicitly the summation over $k_{1,0}$ and $k_{2,0}$ in each
case.

We first begin with the case~:
\begin{equation}
\left \{
\begin{array}{ccc}
\cos (\frac{\pi}{2}k_{1,-}) & > & 0 \\
\cos (\frac{\pi}{2}k_{2,-}) & > & 0 \\
\cos (\frac{\pi}{2}(k_{1,-}+k_{2,-})) & > & 0
\end{array}
\right.   \ ,
\end{equation}
that we will denote as $(+,+,+)$~. The
corresponding contribution to $A_1(T)$ is~:
\begin{multline}
A_1^{(+,+,+)}(T) = -32i\sum_{x_+ \in
\frac{\pi}{2}\mathbb{N}^*} \int dk_{1,0} \, dk_{2,0}
\iint\limits_{\mathcal{T}_{(+,+,+)}} dk_{1,-} \, dk_{2,-}
\\
x_+^2 \frac{e^{\left ( \frac{k_{1,0}}{\pi \cos
    (\frac{\pi}{2}k_{1,-})}+\frac{k_{2,0}}{\pi \cos
    (\frac{\pi}{2}k_{2,-})}-\frac{k_{1,0}+k_{2,0}}{\pi \cos
(\frac{\pi}{2}(k_{1,-}+k_{2,-}))} \right )x_+}}{\cos
(\frac{\pi}{2}k_{1,-})\cos (\frac{\pi}{2}k_{2,-})\cos
(\frac{\pi}{2}(k_{1,-}+k_{2,-}))} \,\chi (k_{1,0} < 0) \,\chi (k_{2,0} <0) \\
\left [ e^{\frac{-Tx_+}{\cos (\frac{\pi}{2}(k_{1,-}+k_{2,-}))}}\chi
  (k_{1,0}+k_{2,0} > -\pi T) 
 - e^{\frac{Tx_+}{\cos (\frac{\pi}{2}(k_{1,-}+k_{2,-}))}}\chi
  (k_{1,0}+k_{2,0} > \pi T ) \right ]\ ,
\end{multline}
where $\mathcal{T}_{(+,+,+)}$ denotes the subset of $\mathcal{T}$
where the signs of the cosines are $(+,+,+)$ respectively. Since the conditions $k_{1,0} < 0$,
$k_{2,0} < 0$ and $k_{1,0}+k_{2,0} > \pm \pi T$ are incompatible,
$A_1^{(+,+,+)} = 0$ .

\subsubsection{The $(+,+,-)$ case}
Let us consider the case~:
\begin{equation}
\left \{
\begin{array}{ccc}
\cos (\frac{\pi}{2}k_{1,-}) & > & 0 \\
\cos (\frac{\pi}{2}k_{2,-}) & > & 0 \\
\cos (\frac{\pi}{2}(k_{1,-}+k_{2,-})) & < & 0
\end{array}
\right. \ ,
\end{equation}
corresponding to the sign configuration $(+,+,-)$. We have~:
\begin{multline}
A_1^{(+,+,-)}(T) = -32i\sum_{x_+ \in
  \frac{\pi}{2}\mathbb{N}^*} \int dk_{1,0} \, dk_{2,0} 
  \iint\limits_{\mathcal{T}_{(+,+,-)}} dk_{1,-} \, dk_{2,-}
  \\
x_+^2 \frac{e^{\left ( \frac{k_{1,0}}{\pi \cos
    (\frac{\pi}{2}k_{1,-})}+\frac{k_{2,0}}{\pi \cos
    (\frac{\pi}{2}k_{2,-})}-\frac{k_{1,0}+k_{2,0}}{\pi \cos
    (\frac{\pi}{2}(k_{1,-}+k_{2,-}))} \right )x_+}}{\cos
  (\frac{\pi}{2}k_{1,-})\cos (\frac{\pi}{2}k_{2,-})\cos
(\frac{\pi}{2}(k_{1,-}+k_{2,-}))} \,\chi (k_{1,0} < 0) \,\chi (k_{2,0} < 
 0) \\
\left [ e^{\frac{-Tx_+}{\cos (\frac{\pi}{2}(k_{1,-}+k_{2,-}))}}\chi
  (k_{1,0}+k_{2,0} < -\pi T) 
 - e^{\frac{Tx_+}{\cos (\frac{\pi}{2}(k_{1,-}+k_{2,-}))}}\chi
  (k_{1,0}+k_{2,0} < \pi T ) \right ] \ .
\end{multline}
The conditions $\chi (k_{1,0}+k_{2,0} < \pm \pi T)$ can obviously be
omitted. We must compute the following expression~:
\begin{multline}
\sum_{n=0}^{+\infty}\sum_{p=0}^{+\infty} e^{-(2n+1)\left (
    \frac{1}{\cos (\frac{\pi}{2}k_{1,-})}-\frac{1}{\cos
      (\frac{\pi}{2}k_{1,-}+k_{2,-})} \right )Tx_+}
e^{-(2p+1)\left ( \frac{1}{\cos (\frac{\pi}{2}k_{2,-})}-\frac{1}{\cos
    (\frac{\pi}{2}k_{1,-}+k_{2,-})} \right )Tx_+} \\
\left [ e^{\frac{-Tx_+}{\cos (\frac{\pi}{2}(k_{1,-}+k_{2,-}))}} -
    e^{\frac{Tx_+}{\cos (\frac{\pi}{2}(k_{1,-}+k_{2,-}))}} \right ]\ ,
\end{multline}
which gives~:
\begin{equation}
\frac{e^{- \left ( \frac{1}{\cos (\frac{\pi}{2}k_{1,-})}+\frac{1}{\cos
        (\frac{\pi}{2}k_{2,-})}-\frac{1}{\cos
        (\frac{\pi}{2}k_{1,-}+k_{2,-})} \right )Tx_+} \left [
    1-e^{\frac{2Tx_+}{\cos (\frac{\pi}{2}(k_{1,-}+k_{2,-}))}} \right ]}
{\left [ 1-e^{-2 \left ( \frac{1}{\cos (\frac{\pi}{2}k_{1,-})} -
        \frac{1}{\cos (\frac{\pi}{2}k_{1,-}+k_{2,-})} \right )Tx_+}
  \right ] \left [ 1-e^{-2 \left ( \frac{1}{\cos
(\frac{\pi}{2}k_{2,-})} - \frac{1}{\cos (\frac{\pi}{2}k_{1,-}+k_{2,-})} \right )Tx_+} \right ]}\ .
\end{equation}
This is clearly a positive real number, and therefore we conclude that 
\begin{equation}
i A_1^{(+,+,-)}(T) \leq 0 \ .
\end{equation}
Indeed, the minus sign of the prefactor $-32i$ is compensated by the
minus sign of the product $\cos (\frac{\pi}{2}k_{1,-})\cos (\frac{\pi}{2}k_{2,-})\cos (\frac{\pi}{2}k_{1,-}+k_{2,-})$.

\subsubsection{The $(+,-,+)$ case}

We now consider the $(+,-,+)$ case. The corresponding contribution
writes~:
\begin{multline}
A_1^{(+,-,+)}(T) = -32i\sum_{x_+ \in
\frac{\pi}{2}\mathbb{N}^*} \int dk_{1,0} \, dk_{2,0} 
  \iint\limits_{\mathcal{T}_{(+,-,+)}} dk_{1,-} \, dk_{2,-}
  \\
x_+^2 \frac{e^{\left ( \frac{k_{1,0}}{\pi \cos
    (\frac{\pi}{2}k_{1,-})}+\frac{k_{2,0}}{\pi \cos
    (\frac{\pi}{2}k_{2,-})}-\frac{k_{1,0}+k_{2,0}}{\pi \cos
    (\frac{\pi}{2}(k_{1,-}+k_{2,-}))} \right )x_+}}{\cos
  (\frac{\pi}{2}k_{1,-})\cos (\frac{\pi}{2}k_{2,-})\cos
(\frac{\pi}{2}(k_{1,-}+k_{2,-}))} \,\chi (k_{1,0} < 0) \,\chi (k_{2,0} > 
 0) \\
\left [ e^{\frac{-Tx_+}{\cos (\frac{\pi}{2}(k_{1,-}+k_{2,-}))}}\chi
  (k_{1,0}+k_{2,0} > -\pi T) 
 - e^{\frac{Tx_+}{\cos (\frac{\pi}{2}(k_{1,-}+k_{2,-}))}}\chi
  (k_{1,0}+k_{2,0} > \pi T ) \right ] \ .
\end{multline}
Here like in all the other cases, we have to sum geometric sequences whose ratio is explicitly strictly smaller than
1. This facilitates the discussion of the
signs of the corresponding quantities, as we shall see. If we perform 
the summation over $k_{1,0}$, we
are lead to a geometric sequence whose ratio is 
$e^{-2 \ \left ( \frac{1}{\cos (\frac{\pi}{2}k_{1,-})} - \frac{1}{\cos
(\frac{\pi}{2}k_{1,-}+k_{2,-})} \right )}$, which will lead to a
factor $\bigl[ 1-e^{-2 \ \left ( \frac{1}{\cos
(\frac{\pi}{2}k_{1,-})} - \frac{1}{\cos
(\frac{\pi}{2}k_{1,-}+k_{2,-})} \right )}  \bigr]^{-1}$ whose sign is not
uniform in $(k_{1,-},k_{2,-})$.

Consequently we introduce the variable $s=k_{1,0}+k_{2,0}$ and replace
$k_{2,0}$ by $s-k_{1,0}$. We must compute~:
\begin{multline}
\int dk_{1,0} \, ds \,e^{\left ( \frac{k_{1,0}}{\pi \cos
(\frac{\pi}{2}k_{1,-})}+\frac{s-k_{1,0}}{\pi \cos
(\frac{\pi}{2}k_{2,-})}-\frac{s}{\pi \cos
(\frac{\pi}{2}(k_{1,-}+k_{2,-}))} \right )x_+} \chi (k_{1,0} < 0)
\chi (s > k_{1,0}) \\ 
\left [ e^{\frac{-Tx_+}{\cos (\frac{\pi}{2}(k_{1,-}+k_{2,-}))}} \chi
(s > -\pi T) - e^{\frac{Tx_+}{\cos
(\frac{\pi}{2}(k_{1,-}+k_{2,-}))}} \chi ( s > \pi T) \right ] \ .
\end{multline}

The variable $s$ describes the set $2\pi T \mathbb{Z}$ and the
condition $\chi ( s > k_{1,0})$ can be omitted. Thus the previous
expression writes~:
\begin{multline}
\sum_{n=0}^{+\infty}e^{-(2n+1)\left ( \frac{1}{\cos
(\frac{\pi}{2}k_{1,-})} - \frac{1}{\cos (\frac{\pi}{2}k_{2,-})}\right ) } 
\left [ e^{\frac{-Tx_+}{\cos (\frac{\pi}{2}(k_{1,-}+k_{2,-}))}}
\sum_{p=0}^{+\infty} e^{-2p \left ( \frac{1}{\cos
(\frac{\pi}{2}(k_{1,-}+k_{2,-}))} - \frac{1}{\cos
(\frac{\pi}{2}k_{2,-})} \right )Tx_+} \right.  
\\
- \left. e^{\frac{-Tx_+}{\cos (\frac{\pi}{2}(k_{1,-}+k_{2,-}))}}
\sum_{p=0}^{+\infty} e^{-2p \left ( \frac{1}{\cos
(\frac{\pi}{2}(k_{1,-}+k_{2,-}))} - \frac{1}{\cos
(\frac{\pi}{2}k_{2,-})} \right )Tx_+} \right ]
\end{multline}
which is equal to~:
\begin{equation}
\frac{e^{- \left ( \frac{1}{\cos (\frac{\pi}{2}k_{1,-})}-\frac{1}{\cos
        (\frac{\pi}{2}k_{2,-})}+\frac{1}{\cos
        (\frac{\pi}{2}(k_{1,-}+k_{2,-}))} \right )Tx_+} \left [
    1-e^{\frac{2Tx_+}{\cos (\frac{\pi}{2}k_{2,-})}} \right ]}
{\left [ 1-e^{-2 \left ( \frac{1}{\cos (\frac{\pi}{2}k_{1,-})} -
        \frac{1}{\cos (\frac{\pi}{2}k_{2,-})} \right )Tx_+}
  \right ] \left [ 1-e^{-2 \left ( \frac{1}{\cos
          (\frac{\pi}{2}(k_{1,-}+k_{2,-}))} - \frac{1}{\cos
          (\frac{\pi}{2}k_{2,-})} \right )Tx_+} \right ]}\ .
\end{equation}
This quantity is positive, thus the conclusion follows~:
\begin{equation}
i A_1^{(+,-,+)}(T) \leq 0\ .
\end{equation}

\subsubsection{The $(+,-,-)$ case}

Let us examine now the $(+,-,-)$ case. The contribution is~:
\begin{multline}
A_1^{(+,-,-)}(T) = -32i\sum_{x_+ \in
  \frac{\pi}{2}\mathbb{N}^*} \int dk_{1,0} \, dk_{2,0} 
  \iint\limits_{\mathcal{T}_{(+,-,-)}} dk_{1,-} \, dk_{2,-}
  \\
x_+^2 \frac{e^{\left ( \frac{k_{1,0}}{\pi \cos
    (\frac{\pi}{2}k_{1,-})}+\frac{k_{2,0}}{\pi \cos
    (\frac{\pi}{2}k_{2,-})}-\frac{k_{1,0}+k_{2,0}}{\pi \cos
    (\frac{\pi}{2}(k_{1,-}+k_{2,-}))} \right )x_+}}{\cos
  (\frac{\pi}{2}k_{1,-})\cos (\frac{\pi}{2}k_{2,-})\cos
  (\frac{\pi}{2}(k_{1,-}+k_{2,-}))} \,\chi (k_{1,0} < 0) \,\chi (k_{2,0} > 
  0) \\
\left [ e^{\frac{-Tx_+}{\cos (\frac{\pi}{2}(k_{1,-}+k_{2,-}))}}\chi
  (k_{1,0}+k_{2,0} < -\pi T) 
 - e^{\frac{Tx_+}{\cos (\frac{\pi}{2}(k_{1,-}+k_{2,-}))}}\chi
  (k_{1,0}+k_{2,0} < \pi T ) \right ] \ .
\end{multline}
We set $k_{1,0}=s-k_{2,0}$ and we compute~:
\begin{multline}
\int ds  \, dk_{2,0} \,e^{\left ( \frac{s-k_{2,0}}{\pi \cos
    (\frac{\pi}{2}k_{1,-})}+\frac{k_{2,0}}{\pi \cos
    (\frac{\pi}{2}k_{2,-})}-\frac{s}{\pi \cos
    (\frac{\pi}{2}(k_{1,-}+k_{2,-}))} \right )x_+} \chi (s < k_{2,0})
    \chi (k_{2,0} > 0) \\ 
\left [ e^{\frac{-Tx_+}{\cos (\frac{\pi}{2}(k_{1,-}+k_{2,-}))}} \chi
    (s < -\pi T) - e^{\frac{Tx_+}{\cos
    (\frac{\pi}{2}(k_{1,-}+k_{2,-}))}} \chi ( s < \pi T) \right ]\ .
\end{multline}
The condition $\chi (s < k_{2,0})$ may be omitted and we must
evaluate~:
\begin{multline}
\sum_{n=0}^{+\infty} e^{(2n+1) \left ( \frac{1}{\cos
      (\frac{\pi}{2}k_{2,-})}-\frac{1}{\cos (\frac{\pi}{2}k_{1,-})}
      \right ) Tx_+} \\
\left [ e^{\frac{-Tx_+}{\cos (\frac{\pi}{2}(k_{1,-}+k_{2,-}))}}
      \sum_{p=1}^{+\infty}e^{-2p \left ( \frac{1}{\cos
      (\frac{\pi}{2}k_{1,-})} - \frac{1}{\cos
      (\frac{\pi}{2}(k_{1,-}+k_{2,-}))} \right ) Tx_+} \right. \\
- \left. e^{\frac{Tx_+}{\cos (\frac{\pi}{2}(k_{1,-}+k_{2,-}))}}
      \sum_{p=0}^{+\infty}e^{-2p \left ( \frac{1}{\cos
      (\frac{\pi}{2}k_{1,-})} - \frac{1}{\cos
      (\frac{\pi}{2}(k_{1,-}+k_{2,-}))} \right ) Tx_+} \right ] \ .
\end{multline}
We find~:
\begin{equation}
\frac{e^{-\left ( \frac{1}{\cos (\frac{\pi}{2}k_{1,-})} -
      \frac{1}{\cos (\frac{\pi}{2}k_{2,-})} - \frac{1}{\cos
        (\frac{\pi}{2}(k_{1,-}+k_{2,-}))} \right ) Tx_+} \left [
    e^{\frac{-2Tx_+}{\cos (\frac{\pi}{2}k_{1,-})}} - 1 \right ]}
{\left [ 1- e^{-2 \left ( \frac{1}{\cos (\frac{\pi}{2}k_{1,-})} -
        \frac{1}{\cos (\frac{\pi}{2}k_{2,-})} \right ) Tx_+} \right ]
\left [ 1- e^{-2 \left ( \frac{1}{\cos (\frac{\pi}{2}k_{1,-})} -
\frac{1}{\cos (\frac{\pi}{2}(k_{1,-}+k_{2,-}))} \right ) Tx_+}
\right ]} \ .
\end{equation}
This is a negative number, therefore 
\begin{equation}
i A_1^{(+,-,-)}(T) \leq 0 \ .
\end{equation}

\subsubsection{The $(-,+,+)$ and (-,+,-) cases}

There is no discussion to carry out~: in fact, for $(k_{1,-},k_{2,-})
\in \mathcal{T}$, we have never $\cos (\frac{\pi}{2}k_{1,-}) < 0$,
$\cos (\frac{\pi}{2}k_{2,-}) > 0$ and $\cos
(\frac{\pi}{2}(k_{1,-}+k_{2,-})) < 0$ simultaneously.
We also conclude in the same way for the $(-,+,-)$ case.

\subsubsection{The (-,-,+) case}

\begin{multline}
A_1^{(-,-,+)}(T) = -32i\sum_{x_+ \in
  \frac{\pi}{2}\mathbb{N}^*} \int dk_{1,0} \, dk_{2,0} 
  \iint\limits_{\mathcal{T}_{(-,-,+)}} dk_{1,-} \, dk_{2,-}
  \\
x_+^2 \frac{e^{\left ( \frac{k_{1,0}}{\pi \cos
    (\frac{\pi}{2}k_{1,-})}+\frac{k_{2,0}}{\pi \cos
    (\frac{\pi}{2}k_{2,-})}-\frac{k_{1,0}+k_{2,0}}{\pi \cos
    (\frac{\pi}{2}(k_{1,-}+k_{2,-}))} \right )x_+}}{\cos
  (\frac{\pi}{2}k_{1,-})\cos (\frac{\pi}{2}k_{2,-})\cos
  (\frac{\pi}{2}(k_{1,-}+k_{2,-}))} \,\chi (k_{1,0} > 0) \,\chi (k_{2,0} > 
  0) \\
\left [ e^{\frac{-Tx_+}{\cos (\frac{\pi}{2}(k_{1,-}+k_{2,-}))}}\chi
  (k_{1,0}+k_{2,0} > -\pi T) 
 - e^{\frac{Tx_+}{\cos (\frac{\pi}{2}(k_{1,-}+k_{2,-}))}}\chi
  (k_{1,0}+k_{2,0} > \pi T ) \right ]\ .
\end{multline}
We remark that the conditions $\chi ( k_{1,0}+k_{2,0} > \pm \pi T)$
are superfluous, and that there is no need to introduce the variable
$s$. We have~:
\begin{multline}
\sum_{n=0}^{+\infty} e^{(2n+1) \left (\frac{1}{\cos
      (\frac{\pi}{2}k_{1,-})} - \frac{1}{\cos
      (\frac{\pi}{2}(k_{1,-}+k_{2,-}))} \right ) Tx_+} 
\sum_{p=0}^{+\infty} e^{(2p+1) \left (\frac{1}{\cos
      (\frac{\pi}{2}k_{2,-})} - \frac{1}{\cos
      (\frac{\pi}{2}(k_{1,-}+k_{2,-}))} \right ) Tx_+} \\
\left [ e^{\frac{-Tx_+}{\cos (\frac{\pi}{2}(k_{1,-}+k_{2,-}))}} -
      e^{\frac{Tx_+}{\cos (\frac{\pi}{2}(k_{1,-}+k_{2,-}))}} \right ]
      = \\
\frac{e^{-\left ( \frac{1}{\cos (\frac{\pi}{2}(k_{1,-}+k_{2,-}))} -
      \frac{1}{\cos (\frac{\pi}{2}k_{1,-})} - \frac{1}{\cos
        (\frac{\pi}{2}k_{2,-})} \right ) Tx_+} \left [
    e^{\frac{-2Tx_+}{\cos (\frac{\pi}{2}(k_{1,-}+k_{2,-}))}} - 1 \right ]}
{\left [ 1- e^{-2 \left ( \frac{1}{\cos (\frac{\pi}{2}(k_{1,-}+k_{2,-}))} -
        \frac{1}{\cos (\frac{\pi}{2}k_{1,-})} \right ) Tx_+} \right ]
\left [ 1- e^{-2 \left ( \frac{1}{\cos (\frac{\pi}{2}(k_{1,-}+k_{2,-}))} -
        \frac{1}{\cos (\frac{\pi}{2}k_{2,-})} \right ) Tx_+} \right ]} \ .
\end{multline}

This quantity is negative and we conclude that 
\begin{equation}
i A_1^{(-,-,+)}(T) \leq 0 \ .
\end{equation}

\subsubsection{The $(-,-,-)$ case}

We finally discuss the last case~: 
\begin{multline}
A_1^{(-,-,-)}(T) = -32i\sum_{x_+ \in
  \frac{\pi}{2}\mathbb{N}^*} \int dk_{1,0} \, dk_{2,0} 
  \iint\limits_{\mathcal{T}_{(-,-,-)}} dk_{1,-} \, dk_{2,-}
  \\
x_+^2 \frac{e^{\left ( \frac{k_{1,0}}{\pi \cos
    (\frac{\pi}{2}k_{1,-})}+\frac{k_{2,0}}{\pi \cos
    (\frac{\pi}{2}k_{2,-})}-\frac{k_{1,0}+k_{2,0}}{\pi \cos
    (\frac{\pi}{2}(k_{1,-}+k_{2,-}))} \right )x_+}}{\cos
  (\frac{\pi}{2}k_{1,-})\cos (\frac{\pi}{2}k_{2,-})\cos
  (\frac{\pi}{2}(k_{1,-}+k_{2,-}))} \,\chi (k_{1,0} > 0) \,\chi (k_{2,0} > 
  0) \\
\left [ e^{\frac{-Tx_+}{\cos (\frac{\pi}{2}(k_{1,-}+k_{2,-}))}}\chi
  (k_{1,0}+k_{2,0} < -\pi T) 
 - e^{\frac{Tx_+}{\cos (\frac{\pi}{2}(k_{1,-}+k_{2,-}))}}\chi
  (k_{1,0}+k_{2,0} < \pi T ) \right ]\ .
\end{multline}
But it is clear that the conditions $k_{1,0} > 0$, $k_{2,0} > 0$ and
$k_{1,0}+k_{2,0} < \pm \pi T$ are incompatible (as in the $(+,+,+)$
case), hence 
\begin{equation}
A_1^{(-,-,-)}(T) = 0 \ .
\end{equation}

\begin{lemma}
There exists a constant $K > 0$ such that~:
\begin{equation}
\left| A_1^{(+,+,-)} (T) + A_1^{(+,-, +)} (T) 
+ A_1^{(+, -, -)} (T) + A_1^{(-, -, +)} (T) \right| > 
\frac{K}{T} \ .
\end{equation}
\end{lemma}

\medskip\noindent
\textbf{Proof~:}\ 
As each one of the quantities are purely imaginary, with non-negative
imaginary part, it is sufficient to prove the inequality $|A_1^{(+, +, -)}
(T)| > \frac{K_1}{T}$ for some constant $K_1$. We have~:
\begin{multline}
|A_1^{(+, +, -)} (T)| = 32 \sum_{x_+ \in \frac{\pi}{2} \mathbb{N}^*}
 \int\int_{\mathcal{T}^{(+, +, -)}} dk_{1, -} dk_{2, -} \\
x_+^2 \frac{e^{-\left( \frac{1}{\cos \frac{\pi}{2} k_{1, -}} + \frac{1}{\cos
 \frac{\pi}{2} k_{2, -}} - \frac{1}{\cos \frac{\pi}{2} (k_{1, -} + k_{2,
 -})}\right) T x_+} \left[ 1- e^{\frac{2 T x_+}{\cos \frac{\pi}{2} (k_{1, 
 -} +
 k_{2, -})}} \right]}{\left[ 1-e^{-2 \left( \frac{1}{\cos \frac{\pi}{2}
 k_{1,-}} - \frac{1}{\cos \frac{\pi}{2} (k_{1, -} + k_{2, -})} \right) T 
 x_+}
 \right] \left[ 1-e^{-2 \left( \frac{1}{\cos \frac{\pi}{2}
 k_{2,-}} - \frac{1}{\cos \frac{\pi}{2} (k_{1, -} + k_{2, -})} \right) T 
 x_+}
 \right]  }\ .
\end{multline}
As $\left[ 1-e^{-2 \left( \frac{1}{\cos \frac{\pi}{2}
k_{1,-}} - \frac{1}{\cos \frac{\pi}{2} (k_{1, -} + k_{2, -})} \right) T 
x_+} \right] \leq 1$ and $\left[ 1-e^{-2 \left( \frac{1}{\cos 
\frac{\pi}{2} k_{2,-}} - 
\frac{1}{\cos \frac{\pi}{2} (k_{1, -} + k_{2, -})} \right) 
T x_+} \right] \leq 1$, we get~:
\begin{multline}
|A_1^{(+, +, -)} (T)| = 32 \sum_{x_+ \in \frac{\pi}{2} \mathbb{N}^*}
 \int\int_{\mathcal{T}^{(+, +, -)}} dk_{1, -} dk_{2, -} \\
x_+^2 e^{-\left( \frac{1}{\cos \frac{\pi}{2} k_{1, -}} + \frac{1}{\cos
 \frac{\pi}{2} k_{2, -}} - \frac{1}{\cos \frac{\pi}{2} (k_{1, -} + k_{2,
 -})}\right) T x_+} \left[ 1- e^{\frac{2 T x_+}{\cos \frac{\pi}{2} (k_{1, 
 -} +  k_{2, -})}} \right] \ .
\end{multline}

As we are seeking a lower bound, we can restrict the integration over the open domain $\mathcal{T}^{(+, +, -)}$ to a compact $\mathcal{T}_{\epsilon}^{(+, +, -)} \subset \mathcal{T}^{(+, +, -)}$, where $\epsilon$ is a strictly
positive constant (for example $\epsilon = \frac{1}{10}$), in
which we have $\cos \left( \frac{\pi}{2} k_{1, -} \right) \geq \epsilon$,
$\cos \left( \frac{\pi}{2} k_{2, -} \right) \geq \epsilon$ and $\left| \cos
\left( \frac{\pi}{2} (k_{1, -} + k_{2, -}) \right) \right| \geq
\epsilon$. 
For $(k_{1, -}, k_{2, -}) \in \mathcal{T}_\epsilon^{(+, +, -)}$, we have~:
\begin{equation}
0 < x_+^2. e^{-\left( \frac{1}{\cos \frac{\pi}{2} k_{1, -}} + \frac{1}{\cos
 \frac{\pi}{2} k_{2, -}} - \frac{1}{\cos \frac{\pi}{2} (k_{1, -} + k_{2,
 -})}\right) T x_+} \left[ 1- e^{\frac{2 T x_+}{\cos \frac{\pi}{2} (k_{1,  
 -} + k_{2, -})}} \right] \leq e^{- 3 T x_+} \ .
\end{equation}
By Lebesgue domination theorem, we can invert $\sum_{x_+}$ and $\int
\int_{\mathcal{T}_\epsilon^{(+, +, -)}} dk_{1,-} \ dk_{2, -}$ and write~:
\begin{equation}
|A_1^{(+, +, -)} (T)| \geq 32 \int_{\mathcal{T}_\epsilon^{(+, +, -)}} dk_{1,-} \ dk_{2, -} \sum_{x_+ \in \frac{\pi}{2} \mathbb{N}^*} x_+^2
 . e^{-\frac{3}{\epsilon} T x_+} \left[ 1 - e^{- 2 T x_+} \right]\ ,
\end{equation}
or~:
\begin{equation}
|A_1^{(+, +, -)} (T)| \geq 8 \pi^2 \sum_{n = 0}^{+ \infty} n^2 e^{- \frac{3
 \pi}{2 \epsilon} T n} \left( 1 - e^{- \pi T n} \right) \ .
\end{equation}
Now we use the formula~:
\begin{equation}
\sum_{n = 0}^{+ \infty} n^2 e^{- a n} = \frac{e^{- a} + e^{- 2 a}}{(1 - e^{- a})^3} \ (\text{for} \ a > 0)
\end{equation}
to write~:
\begin{eqnarray}
|A_1^{(+, +, -)} (T)| & \geq & 8 \pi^2 \left( \frac{e^{- \frac{3 \pi}{2 \epsilon}
 T} + e^{- \frac{3 \pi}{\epsilon} T}}{\left( 1 - e^{- \frac{3 \pi}{2 
 \epsilon}
 T} \right)^3 } - \frac{e^{- \left( \frac{3 \pi}{2 \epsilon} + 1 \right) 
 T} +
 e^{- \left(\frac{3 \pi}{\epsilon} + 2  \right) T}}{ \left( 1 - e^{- \left(
 \frac{3 \pi}{\epsilon} + 2 \right) T} \right)^3} \right) \\
& \geq & 8 \pi^2 \left( \frac{e^{- \frac{3 \pi}{2 \epsilon}
 T} + e^{- \frac{3 \pi}{\epsilon} T} - e^{- \left( \frac{3 \pi}{2 
 \epsilon} +
 1 \right) T} - e^{- \left(\frac{3 \pi}{\epsilon} + 2  \right) T} }{\left( 
 1 - e^{- \frac{3 \pi}{2 \epsilon}T} \right)^3  } \right)\ .
\end{eqnarray}

Using the inequality $\frac{1}{1 - e^{- \frac{3 \pi}{2 \epsilon}} T} \geq
\frac{2 \epsilon}{3 \pi T}$ and assuming that $T < 1$, we obtain the desired
result~:
\begin{equation}
|A_1^{(+, +, -)} (T)| \geq 8 \pi^2 \frac{(2 \epsilon)^3}{(3 \pi)^3} \frac{e^{-
 \frac{3 \pi}{2 \epsilon}} (1 - e^{-1}) + e^{- \frac{3 \pi}{\epsilon}} (1 -
 e^{-2})}{T^3} \ .
\end{equation}
Therefore the lemma is proven.

\section{Study of the other configurations}

We now are going to treat the other configuration,
corresponding to~: 
\begin{equation}
\left\{
\begin{array}{ccc}
k_{1,+} & \approx & -1 \\
k_{2,+} & \approx & 1 \\
k_{3,+} & \approx & -1
\end{array}
\right. \ \ 
{\rm and}\ 
\left\{
\begin{array}{ccc}
k_{1,+} & \approx & -1 \\
k_{2,+} & \approx & -1 \\
k_{3,+} & \approx & 1
\end{array}
\right.
\end{equation}
which are equal and form the term called $2\partial _+^2
A_{G,2}(\pi T, 1, 0)$. Let us concentrate on the first case.
We have to consider the propagator~:
\begin{equation}
\int_{-\infty}^{+\infty}dk_{1,+}\frac{e^{ik_{1,+}x_+}}{ik_{1,0}-\pi
  k_{1,+}\cos (\frac{\pi}{2}k_{1,-})} =
\frac{-1}{\pi \cos
(\frac{\pi}{2}k_{1,-})}\int_{-\infty}^{+\infty}dk_{1,+}\frac{e^{ik_{1,+}x
  _+}}{k_{1,+}-\frac{ik_{1,0}}{\pi \cos (\frac{\pi}{2}k_{1,-})}}\ .
\end{equation}
The pole of the integrand is $\frac{ik_{1,0}}{\pi \cos
(\frac{\pi}{2}k_{1,-})}$ and the corresponding residue writes
$e^{-\frac{k_{1,0}x_+}{\pi \cos (\frac{\pi}{2}k_{1,-})}}$.
Therefore we have~:
\begin{multline}
\int_{-\infty}^{+\infty}dk_{1,+}\frac{e^{ik_{1,+}x_+}}{ik_{1,0}-\pi k_{1,+}\cos (\frac{\pi}{2}k_{1,-})} =
\frac{-2i}{\cos (\frac{\pi}{2}k_{1,-})}e^{-\frac{k_{1,0}x_+}{\pi \cos
    (\frac{\pi}{2}k_{1,-})}} \\
\left [\chi (x_+ > 0)\chi \left ( \frac{k_{1,0}}{\pi \cos
      (\frac{\pi}{2}k_{1,-})} > 0 \right ) - \chi (x_+ < 0) \chi \left
    ( \frac{k_{1,0}}{\pi \cos (\frac{\pi}{2}k_{1,-})} < 0 \right
  )\right ]\ .
\end{multline}
Now, let us consider the integration over $k_{2,+}$. We have~:
\begin{equation}
\int_{-\infty}^{+\infty}dk_{2,+}\frac{e^{ik_{2,+}x_+}}{-ik_{2,0}+\pi
  k_{2,+}\cos (\frac{\pi}{2}k_{2,-})} =
\frac{1}{\pi \cos
(\frac{\pi}{2}k_{2,-})}\int_{-\infty}^{+\infty}dk_{2,+}
\frac{e^{ik_{2,+}x_+}}{k_{2,+}-\frac{ik_{2,0}}{\pi \cos (\frac{\pi}{2}k_{2,-})}}\ .
\end{equation}
In fact, the only change with the previous case is a global change of
sign. We can immediately write~:
\begin{multline}
\int_{-\infty}^{+\infty}dk_{2,+}\frac{e^{ik_{2,+}x_+}}{-ik_{2,0}+\pi
  k_{2,+}\cos (\frac{\pi}{2}k_{2,-})} =
\frac{2i}{\cos (\frac{\pi}{2}k_{2,-})}e^{-\frac{k_{2,0}x_+}{\pi \cos
    (\frac{\pi}{2}k_{2,-})}} \\
\left [\chi (x_+ > 0)\chi \left ( \frac{k_{2,0}}{\pi \cos
      (\frac{\pi}{2}k_{2,-})} > 0 \right ) - \chi (x_+ < 0) \chi \left
    ( \frac{k_{2,0}}{\pi \cos (\frac{\pi}{2}k_{2,-})} < 0 \right
  )\right ]\ .
\end{multline}

For the integration over $k_{3,+}$, we have $\cos
(\frac{\pi}{2}k_{2,+}) \approx \frac{\pi}{2}(k_{2,+}+1)$ and we
consider~:
\begin{equation}
\int_{-\infty}^{+\infty}dk_{3,+}\frac{e^{ik_{3,+}x_+}}{-ik_{3,0}-\pi
  k_{3,+}\cos (\frac{\pi}{2}k_{3,-})} = 
\frac{-1}{\pi \cos
(\frac{\pi}{2}k_{3,-})}\int_{-infty}^{+\infty}\frac{e^{ik_{3,+}x_+}}{k_{3,+}+\frac{ik_{3,0}}{\pi \cos(\frac{\pi}{2}k_{3,+})}}\ .
\end{equation}
In this case, the pole is $-\frac{ik_{3,0}}{\pi \cos
(\frac{\pi}{2}k_{3,-})}$ and the residue $e^{\frac{k_{3,0}x_+}{\pi
\cos (\frac{\pi}{2}k_{3,-})}}$. Therefore the above integral
writes~:
\begin{multline}
\int_{-\infty}^{+\infty}dk_{3,+}\frac{e^{ik_{3,+}x_+}}{-ik_{3,0}-\pi
k_{3,+}\cos (\frac{\pi}{2}k_{3,-})} = \frac{-2i}{\cos
(\frac{\pi}{2}k_{3,-})}e^{\frac{k_{3,0}x_+}{\pi \cos
(\frac{\pi}{2}k_{3,-})}} \\
\left [ \chi (x_+ > 0) \chi \left (\frac{k_{3,0}}{\pi \cos
(\frac{\pi}{2}k_{3,-})} < 0 \right ) - \chi (x_+ < 0) \chi \left
(\frac{k_{3,0}}{\pi \cos (\frac{\pi}{2}k_{3,-})} > 0 \right ) \right]\ .
\end{multline}
Hence we obtain~:
\begin{multline}
\partial_+^2 \tilde{A}_{G,2}(\pi T,1,0) = -8i\int dx
\int dk_{1,0} \,dk_{1,-} \,dk_{2,0} \,dk_{2,-} \,dk_{3,0} \,dk_{3,-} \\
x_+^2 \,\frac{e^{\left ( -\frac{k_{1,0}}{\pi \cos(\frac{\pi}{2}k_{1,-})}-\frac{k_{2,0}}{\pi \cos
(\frac{\pi}{2}k_{2,-})}+\frac{k_{3,0}}{\pi \cos
(\frac{\pi}{2}k_{3,-})} \right )x_+}}{\cos
(\frac{\pi}{2}k_{1,-})\cos (\frac{\pi}{2}k_{2,-})\cos
(\frac{\pi}{2}k_{3,-})} \,e^{i(k_{1,0}+k_{2,0}+k_{3,0}+\pi 
T)t}\,e^{i(k_{1,-}+k_{2,-}+k_{3,-})x_-}  \\
\left [ \chi (x_+ > 0) \,\chi \left (
\frac{k_{1,0}}{\pi \cos (\frac{\pi}{2}k_{1,-})} > 0 \right ) \,\chi
\left (
\frac{k_{2,0}}{\pi \cos (\frac{\pi}{2}k_{2,-})} > 0 \right )\chi \left
( \frac{k_{3,0}}{\pi \cos (\frac{\pi}{2}k_{3,-})} < 0 \right ) \right. \\
 - \left.  \chi (x_+ < 0) \,\chi \left (\frac{k_{1,0}}{\pi \cos
       (\frac{\pi}{2}k_{1,-})} < 0 \right ) \,\chi \left
     (\frac{k_{2,0}}{\pi \cos (\frac{\pi}{2}k_{2,-})} < 0 \right )\chi
\left (\frac{k_{3,0}}{\pi \cos (\frac{\pi}{2}k_{3,-})} > 0 \right ) 
\right ]\ .
\end{multline}

Then we integrate over $t$ and perform the sum over $k_{3,0}$~:
\begin{multline}
\partial_+^2 \tilde{A}_{G,2}(\pi T,1,0) = -8i\int dx_+ \, dx_-
\int dk_{1,0} \,dk_{1,-} \,dk_{2,0} \,dk_{2,-} \,dk_{3,-} \\
x_+^2 \,\frac{e^{\left ( -\frac{k_{1,0}}{\pi \cos
(\frac{\pi}{2}k_{1,-})}-\frac{k_{2,0}}{\pi \cos
    (\frac{\pi}{2}k_{2,-})}-\frac{k_{1,0}+k_{2,0}+\pi T}{\pi \cos
    (\frac{\pi}{2}k_{3,-})} \right )x_+}}{\cos
  (\frac{\pi}{2}k_{1,-})\cos (\frac{\pi}{2}k_{2,-})\cos
  (\frac{\pi}{2}k_{3,-})} \,e^{i(k_{1,-}+k_{2,-}+k_{3,-})x_-}  \\
 \left [ \chi (x_+ > 0) \,\chi \left (
\frac{k_{1,0}}{\pi \cos (\frac{\pi}{2}k_{1,-})} > 0 \right ) \,\chi
\left (
\frac{k_{2,0}}{\pi \cos (\frac{\pi}{2}k_{2,-})} > 0 \right )\chi \left
( \frac{k_{3,0}}{\pi \cos (\frac{\pi}{2}k_{3,-})} < 0 \right ) \right. \\
 - \left.  \chi (x_+ < 0) \,\chi \left (\frac{k_{1,0}}{\pi \cos
       (\frac{\pi}{2}k_{1,-})} < 0 \right ) \,\chi \left
     (\frac{k_{2,0}}{\pi \cos (\frac{\pi}{2}k_{2,-})} < 0 \right )\chi
\left (\frac{k_{3,0}}{\pi \cos (\frac{\pi}{2}k_{3,-})} > 0 \right ) 
\right ]\ .
\end{multline}
Thanks to the change of variables $ x_+' =  -x_+ $, 
$k_{1,0}'  =  -k_{1,0}$,
$k_{2,0}'  =  -k_{2,0} $, we get~:
\begin{multline}
\partial_+^2 \tilde{A}_{G,2}(\pi T,1,0) = -8i\int dx_+ \, dx_-
\int dk_{1,0} \,dk_{1,-} \,dk_{2,0} \,dk_{2,-} \,dk_{3,-} \\
x_+^2 \,\frac{e^{-\left ( \frac{k_{1,0}}{\pi \cos
    (\frac{\pi}{2}k_{1,-})}+\frac{k_{2,0}}{\pi \cos
    (\frac{\pi}{2}k_{2,-})}+\frac{k_{1,0}+k_{2,0}}{\pi \cos
    (\frac{\pi}{2}k_{3,-})} \right )x_+}}{\cos
  (\frac{\pi}{2}k_{1,-})\cos (\frac{\pi}{2}k_{2,-})\cos
  (\frac{\pi}{2}k_{3,-})} \,e^{i(k_{1,-}+k_{2,-}+k_{3,-})x_-}  \\
\chi (x_+ > 0) \,\chi \left (
\frac{k_{1,0}}{\pi \cos (\frac{\pi}{2}k_{1,-})} > 0 \right ) \,\chi
\left ( \frac{k_{2,0}}{\pi \cos (\frac{\pi}{2}k_{2,-})} > 0 \right ) \\
\left [ e^{\frac{-Tx_+}{\cos (\frac{\pi}{2}k_{3,-})}} \chi \left (
    \frac{k_{1,0}+k_{2,0}+ \pi T}{\pi \cos (\frac{\pi}{2}k_{3,-})} > 0
  \right ) - e^{\frac{Tx_+}{\cos (\frac{\pi}{2}k_{3,-})}} \chi \left (
    \frac{k_{1,0}+k_{2,0} - \pi T}{\pi \cos (\frac{\pi}{2}k_{3,-})} > 0
  \right ) \right ]\ .
\end{multline}

Then we perform the sum over $x_-$ as
previously and integrate over $k_{3,-}$. There is a
small contribution with a compensating factor $[\chi (x_+ \ even) - \chi (x_+ \ odd)]$ that can be bounded as in Section VI, and
we have again to study the
dominant contribution~:
\begin{multline}
A_2(T) = -8i\sum_{x_+ \in
\frac{\pi}{2}\mathbb{N}^*} \int dk_{1,0} \, dk_{2,0} \, dk_{1,-} \, 
dk_{2,-}
x_+^2 \,\frac{e^{-\left ( \frac{k_{1,0}}{\pi \cos
    (\frac{\pi}{2}k_{1,-})}+\frac{k_{2,0}}{\pi \cos
    (\frac{\pi}{2}k_{2,-})}+\frac{k_{1,0}+k_{2,0}}{\pi \cos
    (\frac{\pi}{2}(k_{1,-}+k_{2,-}))} \right )x_+}}{\cos
  (\frac{\pi}{2}k_{1,-})\cos (\frac{\pi}{2}k_{2,-})\cos
  (\frac{\pi}{2}(k_{1,-}+k_{2,-}))} 
\\ 
\chi \left ( \frac{k_{1,0}}{\pi \cos (\frac{\pi}{2}k_{1,-})} > 0
\right ) \,\chi \left ( \frac{k_{2,0}}{\pi \cos
(\frac{\pi}{2}k_{2,-})} > 0 \right ) 
\biggl[ e^{\frac{-Tx_+}{\cos (\frac{\pi}{2}(k_{1,-}+k_{2,-}))}}\chi
\left ( \frac{k_{1,0}+k_{2,0}+\pi T}{\pi \cos
  (\frac{\pi}{2}(k_{1,-}+k_{2,-}))} > 0 \right ) 
\\ 
- e^{\frac{Tx_+}{\cos (\frac{\pi}{2}(k_{1,-}+k_{2,-}))}}\chi \left
(\frac{k_{1,0}+k_{2,0}-\pi T}{\pi \cos (\frac{\pi}{2}(k_{1,-}+k_{2,-}))} > 0 \right ) \biggr] \ .
\end{multline}

Fortunately, we do not have to carry again a discussion about the
signs of the three cosines. In fact, we can remark that $A_1(T) =
A_2(T)$. To see that, let us perform the following change of variables
in $A_1(T)$~: 
\begin{equation}
\left\{
\begin{array}{ccc}
k_{1,-} & = & k_{1,-}' + 2 \\
k_{2,-} & = & k_{2,-}' + 2
\end{array}
\right. \ ,
\end{equation}
to obtain~:
\begin{multline}
A_2(T) = -8i\sum_{x_+ \in
\frac{\pi}{2}\mathbb{N}^*} \int dk_{1,0}
\, dk_{2,0} \iint\limits_{\mathcal{T'}} dk_{1,-}'  dk_{2,-}' 
x_+^2 \frac{e^{\left ( \frac{k_{1,0}}{\pi \cos
(\frac{\pi}{2}k_{1,-}')}+\frac{k_{2,0}}{\pi \cos
(\frac{\pi}{2}k_{2,-}')}-\frac{k_{1,0}+k_{2,0}}{\pi \cos
    (\frac{\pi}{2}(k_{1,-}'+k_{2,-}'))} \right )x_+}}{\cos
  (\frac{\pi}{2}k_{1,-}')\cos (\frac{\pi}{2}k_{2,-}')\cos
  (\frac{\pi}{2}(k_{1,-}'+k_{2,-}'))} 
\\ 
\chi \left ( \frac{k_{1,0}}{\pi \cos (\frac{\pi}{2}k_{1,-}')} < 0
  \right ) \chi \left ( \frac{k_{2,0}}{\pi \cos
  (\frac{\pi}{2}k_{2,-}')} < 0 \right ) 
\biggl[ e^{\frac{-Tx_+}{\cos (\frac{\pi}{2}(k_{1,-}'+k_{2,-}'))}}\chi
  \left ( \frac{k_{1,0}+k_{2,0}+\pi T}{\pi \cos
  (\frac{\pi}{2}(k_{1,-}'+k_{2,-}'))} > 0 \right )
  \\ 
 - e^{\frac{Tx_+}{\cos (\frac{\pi}{2}(k_{1,-}'+k_{2,-}'))}}\chi \left
(\frac{k_{1,0}+k_{2,0}-\pi T}{\pi \cos (\frac{\pi}{2}(k_{1,-}'+k_{2,-}'))} > 0 \right ) \biggr] \ ,
\end{multline}
where $\mathcal{T'}$ is the triangle $\mathcal{T}$ translated by the
vector $(-2,-2)$. Using the invariance under central symmetry and
translations by vectors of the form $(4n_+,4n_-)$, $(n_+,n_-) \in
\mathbb{Z}^2$, we conclude that $\mathcal{T'}$ may be replaced by
$\mathcal{T}$.

Hence we have proved that $A_1(T) = A_2(T)$. This concludes
the proof of Theorem \ref{thecau} hence of Theorem \ref{main_theorem}.

\end{document}